\documentclass[journal,twocolumn]{IEEEtran}

\usepackage{graphicx}
\usepackage{amsmath}
\usepackage{amssymb}

\usepackage{cite}

\newlength{\FW}
\usepackage[switch]{lineno}
\begin{document}

\title{Cooperative Jamming Detection Using Low-Rank Structure of Received Signal Matrix}

\author{Amir Mehrabian and  Georges Kaddoum,~\IEEEmembership{Senior Member,~IEEE}

\thanks{Amir Mehrabian and Georges Kaddoum are with the LaCIME Laboratory,
Department of Electrical Engineering, École de Technologie Supérieure,
Montreal, QC H3C 0J9, Canada (e-mail: \{amir.mehrabian, georges.kaddoum\}@etsmtl.ca).}}
\markboth{}%
{Shell \MakeLowercase{\textit{et al.}}: Bare Demo of IEEEtran.cls for IEEE Journals}

\maketitle

\begin{abstract}
Wireless communication can be simply subjected to malicious attacks due to its open nature and shared medium. Detecting jamming attacks is the first and necessary step to adopt the anti-jamming strategies.
This paper presents novel cooperative jamming detection methods that use the low-rank structure of the received signal matrix. We employed the likelihood ratio test to propose detectors for various scenarios. We regarded several scenarios with different numbers of friendly and jamming nodes and different levels of available statistical information on noise. We also provided an analytical examination of the false alarm performance of one of the proposed detectors, which can be used to adjust the detection threshold. We discussed the synthetic signal generation and the Monte Carlo (MC)-based threshold setting method, where knowledge of the distribution of the jamming-free signal, as well as several parameters such as noise variance and channel state information (CSI), is required to accurately generate synthetic signals for threshold estimation. Extensive simulations reveal that the proposed detectors outperform several existing methods, offering robust and accurate jamming detection in a collaborative network of sensing nodes.
\end{abstract}


\IEEEpeerreviewmaketitle

\section{Introduction}
For achieving quality wireless communication, it is vital to acquire spectrum awareness in the related environment. Spectrum awareness is crucial in ensuring optimal utilization of available frequencies and mitigating interference. 
Wireless sensor networks (WSNs) can enhance spectrum awareness and have gained interest for their application in the internet of things (IoT), particularly in environmental monitoring.
%Employing a wireless sensor network (WSN) facilitates the acquisition of this awareness with improved accuracy and has attracted many in IoT to monitor the environment and gather information.
One type of awareness can specify whether the sensed channel is idle, or it has been employed by another communicating network. This problem has been modeled as a binary hypothesis testing, called spectrum sensing (SS) problem, in the context of cognitive radio (CR) and has been extensively studied in the literature \cite{Axell}. \par
A more challenging problem is to recognize the type of existing activity in the sensed channel. This is important when malicious entities attempt to disrupt the communication of a network through a variety of jamming attacks. The open and shared nature of wireless channels makes wireless communication susceptible to security threats, particularly jamming attacks initiated by malicious entities \cite{PLS_survey, Kaddoum_Tcom}. In a jamming attack, a malicious entity such as a jamming node (JN) transmits signals intentionally designed to disrupt the communication between legitimate transmitter nodes (TN) and receiver nodes. This disruptive interference can reduce the reliability and efficiency of wireless communication and might completely block the communication. \par

 One prevalent method employed in jamming attacks involves the transmission of high-power signals, flooding the signals of legitimate transmitters. This flooding can lead to the rejection of legitimate signals at the receiver, decreasing the achievable rate \cite{Mughal}.
Furthermore, deceptive jamming attacks represent another strategy adopted by JNs. In this attack, the JN can imitate the signals of legitimate TNs, to deceive receiving nodes into accepting malicious transmissions. This deceptive tactic not only disrupts ongoing communication but also poses a serious threat to the confidentiality and integrity of transmitted information \cite{JAD}.\par
Therefore, reliable jamming attack detection is an essential first step combating the malicious entity and adopting an appropriate anti-jamming strategy. In the literature, several works proposed different methods for jamming detection. In \cite{kurt,Cooperative_WED}, single-input multiple-output (SIMO) and cooperative models for jamming detection have been studied which include both spectrum sensing and jamming detection problems with a quaternary testing problem. The proposed weighted energy detector (ED) in these works has been derived using the likelihood ratio test (LRT). Similarly, in \cite{Razavizadeh}, a jamming detector, based on the LRT for multiple-input multiple-output (MIMO) base station (BS), was derived. \par
Additionally, deceptive attacks, also called spoofing attacks, are hard to combat and severely harmful in the context of pilot contamination. Energy ratio-based methods have been developed using the known pilot signal in the detection process for this type of attack \cite{Enenrgy_ratio_detection,VAR}. The authors of \cite{JAD} employed pilot signals and channel estimation under jamming attacks to detect the activity of the JN based on the energy of noise.\par
In contrast to the energy-based detection, several works in the literature exploit the low-rank structure and eigenvalues of the received signal for detection. In CR, for spectrum sensing, various works used eigenvalues 
\cite{Taherpour,Mehrabian,Mehrabian_SSJ}, while in jamming detection, eigenvalues and rank of the received matrix were employed for determining the number of signal sources \cite{Vinogradova,FDC_source} by computing metrics, such as the minimum length description (MDL) \cite{MDL,Tugnait_MDL,Nayebi} and Akaike information criterion (AIC) \cite{AIC}. However,  in contrast to LRT-based detectors, these methods lack a detection threshold, and consequently, lack the ability to adjust the probability of false alarm. \par
Due to the variation in the rank of the received matrices, analysis of eigenvalues based on the random matrix theory (RMT) can help detect the active signal sources. In the RMT literature, numerous works have studied the statistical behavior of the eigenvalues of large random matrices, such as noise matrices, in wireless systems \cite{Couillet, SILVERSTEIN1995331}. In \cite{Zanella2003, Zanella2009}, the authors studied the exact joint distribution of eigenvalues of central and non-central matrices with Wishart distribution, and they applied these results to the analysis of MIMO systems. For large matrices with Wishart distribution, using RMT, it is possible to approximate the distribution of the largest eigenvalue with a Tracy-Widom (TW) distribution \cite{Couillet}.  Based on this, the TW distribution was used for the estimation of the number of JNs in  \cite{nadler, RMT_tugnait}, which is referred to as the RMT estimator. However, given the sensitivity of the detector to variations in the detection threshold, these methods can perform poorly if the detection threshold is not adjusted accurately due to the approximation.\par
In addition to the mentioned methods, numerous works in the literature have proposed feature-based detectors and used features, such as the packet delivery ratio (PDR) or the received signal strength (RSS) \cite{Manet_PDR}, which might cause a delay in the detection of attack \cite{Fast_detec} and pose high computational overheads \cite{HMM}. Several other works also use machine learning (ML) methods. In \cite{Wifi}, a protocol using deep learning was proposed to classify the activity of the sensed channel into three cases, such as the presence of an active jammer, only friendly WiFi signal, or idle channel. In \cite{kurt2}, deep neural networks (DNNs) were trained on signals with orthogonal frequency-division multiplexing (OFDM) to identify the jamming and type of interference. A method for the classification of the type of jamming based on DNNs was proposed in \cite{kaddoum_2}, and a kernelized support vector machine (SVM) integrated with DNN for jamming detection in cellular networks was investigated in \cite{kaddoum_3}. \par 
In the context of  WSNs, the authors of \cite{Dos_attck_wsn} proposed a lightweight detection approach for a denial-of-access attack using the decision tree algorithm. The authors of \cite{WSN-DS} provided a dataset called WSN-DS for the classification of jamming attacks in WSNs and proposed artificial neural networks (ANNs) for the detection of jamming. This dataset was also used in other ML-based studies, e.g., \cite{WSD-nbayse, kaddoum_3}. Federated learning (FL) was proposed in \cite{FL_kaddoum} as a framework for collaborative sensing and detection of jamming attacks, where separate nodes cooperate to offer an effective ML model. The proposed FL approach resulted in an accurate classification of jamming attack types in WSN-DS. 
 It is clear that all the above-mentioned ML-based methods require datasets with large samples to offer an appropriate detection and classification performance, where the generalization ability of these methods is highly dependent on the diversity of the dataset. \par
%The survey in \cite{survey_jamming} provides a very detailed review of different existing jamming detection approaches as well as anti-jamming strategies for various conditions and networks.\par
Motivated by this fact, we provide an LRT-based jamming detection method with no need for a prior dataset. In this paper, we focus on a cooperative network while few studies consider cooperative jamming detection methods. In this system model, sensing nodes (SNs) work collaboratively to detect jamming in the fusion center (FC). This enables us to model the received signal with a low-rank structure and derive LRT-based detectors for jamming detection. Detecting deceptive jamming attacks is challenging since the jamming signal closely resembles the desired signal, and it might cause serious damage to the network \cite{survey_jamming}. 
However, by utilizing the low-rank structure, our approach can effectively detect deceptive jamming attacks. 
In comparison with the model-based methods, ML-based detectors demand large prior datasets and a training process with high computational complexity. It is important to note that accessing a dataset that includes information on the jamming signal of malicious nodes is difficult in the real world due to the confidential nature of this data. Malicious jammers intend to hide this information to improve their attack and protect their data from reverse engineering \cite{FL_kaddoum}. In contrast, our approach offers detectors that operate accurately without the need for any prior information on the model's parameters or only require to know the noise variance.
The proposed detectors exploit singular values of the received matrix to achieve high accuracy even for the challenging problem of deceptive jamming detection.\par
In addition, the proposed detectors in this work can serve as a fast initial stage for jamming detection with no need for prior information, addressing the ML-based method's limitations. For benefiting the data-driven methods, if valid datasets are available, and related signals include specific patterns, the ML-based methods can also be integrated with the proposed methods to form a general solution with an effective performance for jamming detection.\par
The main contributions of this work can be listed as follows:
\begin{itemize}
    \item The cooperative scenario enables us to employ the low-rank structure. Then, LRT is derived under the low-rank constraint to derive detection methods according to singular values for various cases conditioned on the availability of statistical information and the number of TNs.
    \item  An analytical examination of the false alarm performance of the derived detectors is provided based on the RMT and using the distribution of eigenvalues of matrices with Wishart distribution. The analytical findings demonstrate strong agreement with simulations, affirming the efficacy of the method for adjusting the detection threshold.
    \item Extensive simulations and various scenarios are presented to compare the performance of the proposed detectors with several existing methods in the literature, confirming the outperformance of the proposed methods.
    \item The effective performance of the proposed detectors is analytically discussed in scenarios where multiple JNs are active. The simulation results also demonstrate that the proposed detectors can effectively detect jamming attacks in the presence of multiple active JNs and TNs.
    \item We also discussed several methods for detection threshold adjustment and their corresponding limitations, including the approach used in this study, which relies on prior knowledge of the jamming-free signal distribution and CSI to accurately generate signals for threshold estimation using the Monte Carlo (MC) method.
    
\end{itemize}
The remaining sections of this paper are structured as follows. The system model is introduced in Section \ref{Ssys}. The proposed methods are presented in  Section \ref{Spropos}. We discuss scenarios with multiple JNs  in Sections \ref{S_MJN}. An analytical discussion on the false alarm probability is provided in Section \ref{Sfalse}. Finally, the simulation results and conclusion of the paper are given in Sections \ref{Simul} and \ref{Scon}, respectively.

\section{System Model}\label{Ssys}
\begin{figure}[tb]
    \centering
    \includegraphics[width =3.5in]{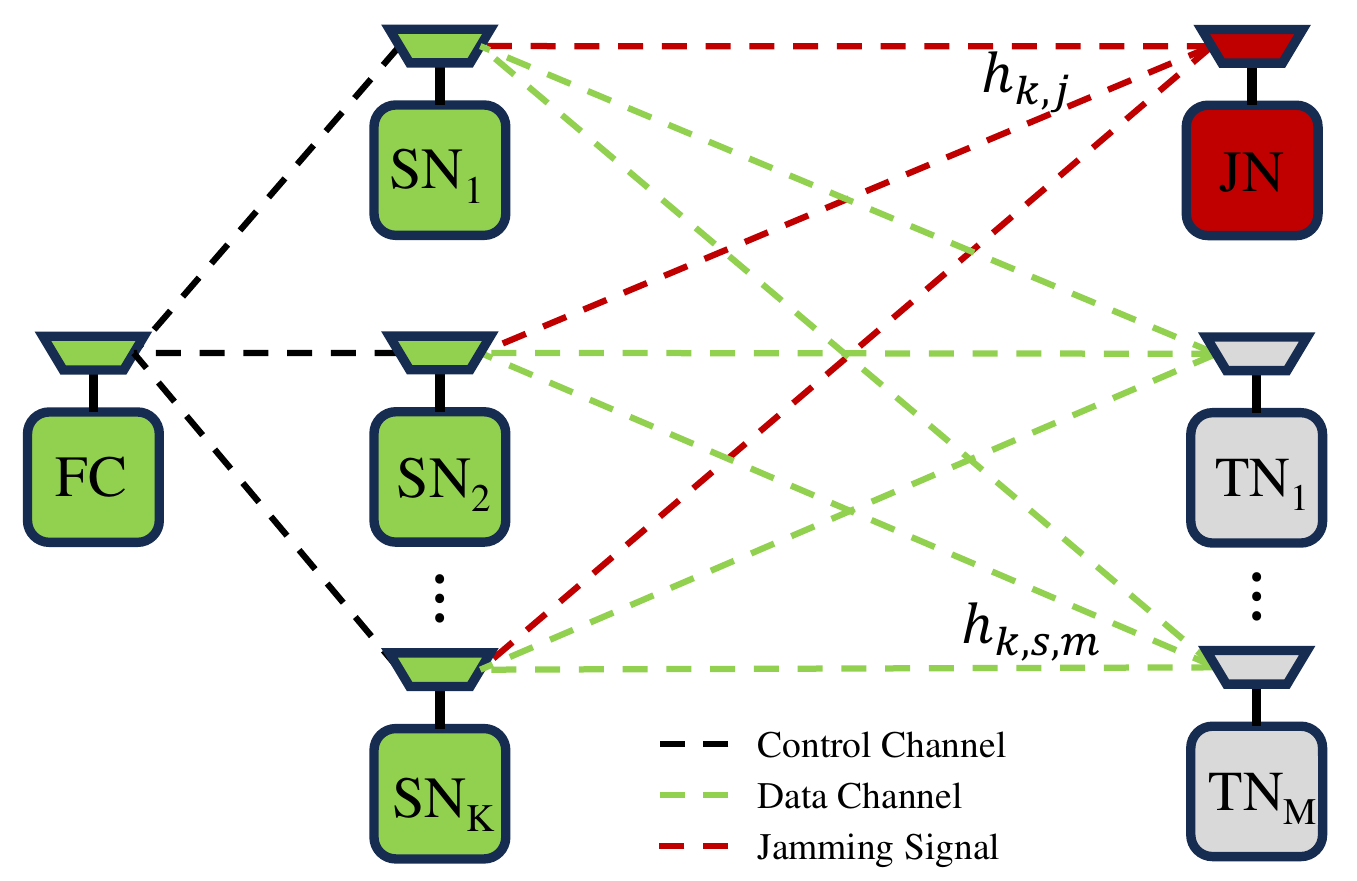}
    \caption{System model for cooperative detection with multiple SNs and TNs and one JN}
    \label{fig0}
\end{figure}
Here, we assume that $K$  SNs are sensing the wireless channel. $M$ TNs are attempting to communicate through the wireless channel with SNs while a malicious JN is transmitting jamming signals to disrupt the friendly communication between TNs and SNs. Fig. \ref{fig0} shows the system model for this network. The received baseband signal in the $k^{th}$ SN can be formulated as
% %%%Therefore, instead of only a single TN, $M$ TNs are active and communicating with the RNs in the network. 
% Since the presence of more TNs increases the rank of $\textbf{Y}$ and changes the assumed low structure of the received matrix, SSV and RSV can not perform properly in this case.\par
% To proceed with $M$ friendly TNs, we rewrite the BHT in (\ref{eq3}) as
% \begin{equation}\label{eq23n}
% \Big{\{}\begin{array}{c}
%      H_0 : \textbf{y}[n] = \sum_{m=1}^M\textbf{h}_{s,m}s_m[n]+\textbf{w}[n] \ \ \ \ \ \ \ \ \ \  \ \\
%      H_1 : \textbf{y}[n] = \sum_{m=1}^M\textbf{h}_{s,m}s_m[n]+\textbf{h}_{j}j[n]+\textbf{w}[n] 
% \end{array},
% \end{equation}
% where $ n = 1, ..., N$, and $\textbf{h}_{s,m}$ is the channel between SNs, and the $m^{th}$ friendly TN and $s_m[n]$ is the transmitted signal by that TN. We  introduce $\textbf{x}_i[n]=\sum_{m=1}^M\textbf{h}_{s,m}s_m[n]$ and $\textbf{X}_i=\sum_{m=1}^M\textbf{h}_{s,m}\textbf{s}^T_m$ to denote vector and matrix of  the received signal before noise contamination under $H_i$, respectively. We, once again, argue that $\textbf{X}_i$ has a low-rank structure and impose this low-rank structure in our solutions using the constraint $C_{M,i}:\mathrm{rank}(\textbf{X}_i)=M+i$.\par
\begin{equation}\label{eq1}
y_k[n] = \sum_{m=1}^Mh_{k,s, m}s_m[n]+h_{k,j}j[n]+w_k[n],
\end{equation}
where $s_m[n]$, $j[n]$, and $w_k[n]$ denote the transmitted signals from the $m^{th}$ TN, from the JN, and noise, respectively, and 
$h_{k,s, m}$ specifies the corresponding channel gain between the $k^{th}$ SN and the $m^{th}$  TN. Similarly, $h_{k,j}$ is the channel gain between the $k^{th}$ SN and the JN. In a cooperative scenario, the FC can collect all the samples acquired in different SNs through the control channel and detect the presence of the JN \cite{Cooperative_WED}. As a result, the collected vector of received SN samples at the FC is 
\begin{equation}\label{eq2}
\textbf{y}[n] = \sum_{m=1}^M\textbf{h}_{s,m}s_m[n]+\textbf{h}_{j}j[n]+\textbf{w}[n],
\end{equation}
where all vectors belong to $\mathbb{C}^{K\times1}$. It is clear that when the JN is off, the received vector at the FC is 
$\textbf{y}[n] = \sum_{m=1}^M\textbf{h}_{s,m}s_m[n]+\textbf{w}[n]$.
During a sensing period, each SN receives $N$ samples. If we assume that channels follow the slow fading model and do not change during one sensing time, we can formulate the jamming detection problem as a binary hypothesis testing (BHT) as follows
\begin{equation}\label{eq3}
\Big{\{}\begin{array}{c}
     H_0 : \textbf{y}[n] =\sum_{m=1}^M\textbf{h}_{s,m}s_m[n]+\textbf{w}[n] \ \ \ \ \ \ \ \ \ \  \ \\
     H_1 : \textbf{y}[n] = \sum_{m=1}^M\textbf{h}_{s,m}s_m[n]+\textbf{h}_{j}j[n]+\textbf{w}[n] 
\end{array}
\end{equation}
where $n = 1, ..., N$. We rewrite it in matrix form as
\begin{equation}\label{eq4}
\Big{\{}\begin{array}{c}
     H_0 : \textbf{Y} =\textbf{X}_0 + \textbf{W} \\
     H_1 : \textbf{Y} = \textbf{X}_1+\textbf{W} 
\end{array},
\end{equation}
in which $H_0$ denotes the case where the JN is off, and $H_1$ is for the case where the JN is on. Moreover, $\textbf{Y}, \textbf{W}, \textbf{X}_0$, and $\textbf{X}_1\in \mathbb{C}^{K\times N}$, and we introduce $\textbf{x}_0[n]=\sum_{m=1}^M\textbf{h}_{s,m}s_m[n]$ and $\textbf{x}_1[n]=\sum_{m=1}^M\textbf{h}_{s,m}s_m[n]+\textbf{h}_{j}j[n]$ to denote the received signal vector before noise contamination under $H_0$ and $H_1$, respectively. The related matrix form can be defined as  $\textbf{X}_0=\sum_{m=1}^M\textbf{h}_{s,m}\textbf{s}^T_m$, and $\textbf{X}_1=\sum_{m=1}^M\textbf{h}_{s,m}\textbf{s}^T_m+\textbf{h}_j\textbf{j}^T$.  The desired signal and jamming signal vectors are $\textbf{s}_m^T = [s_m[1],...,s_m[N]]$, and $\textbf{j}^T = [j[1],...,j[N]]$, respectively, where the superscript $T$ denotes the transposition. Since we assume that the noise samples are uncorrelated both across different SNs and different time samples, and $K<N$, the noise matrix $\textbf{W}$ is a full row-rank matrix with $\mathrm{rank}(\textbf{W})=K$. On the other hand, $\textbf{X}_0$ is created by the sum of the outer product of $\textbf{h}_{s,m}$ and $\textbf{s}_m$, and we assume that the channel vectors $\textbf{h}_{s,m}$ for different TNs are linearly independent, and that the signal vectors $\textbf{s}_m$ for different TNs are also linearly independent. Thus, if $M<K$,  we conclude that $\textbf{X}_0$ has a low-rank structure with $\mathrm{rank}(\textbf{X}_0)=M$. Additionally, throughout this work, we assume that $\textbf{h}_{s,m}$ and $\textbf{h}_j$, along with $\textbf{s}_m$ and $\textbf{j}$ are linearly independent, which implies that $\textbf{X}_1$ is also a low-rank matrix with $\mathrm{rank}(\textbf{X}_1)=M+1$. In the next sections, this low-rank structure of  $\textbf{X}_i$ will be used to design jamming detectors. Importantly, we assume that the channel gains remain fixed during the sensing interval, which enables us to model $\mathbf{X}_i$ as a low-rank matrix. However, in very fast-fading environments, where channel variations are noticeable even within a single sensing period, this low-rank assumption becomes inaccurate, leading to performance degradation of the proposed methods under such conditions.
\par

Here, we consider the case where the noise samples are independent and identically distributed (i.i.d.) with a zero-mean circularly symmetric complex white Gaussian (CSCWG) distribution, $\textbf{w}[n]\sim \mathcal{CN}(\textbf{0},\textbf{R}_w)$, where $\textbf{R}_w$ is the covariance matrix of the noise vector. Then, we can rewrite the BHT with the distribution of $\textbf{y}[n]$ as
\begin{equation}\label{eq5}
\Big{\{}\begin{array}{c}
     H_0 : \textbf{y}[n] \sim \mathcal{CN}(\textbf{x}_{0}[n], \textbf{R}_w)  \\
     H_1 : \textbf{y}[n] \sim \mathcal{CN}(\textbf{x}_{1}[n], \textbf{R}_w) 
\end{array}
\ \ \ \ n = 1, ..., N,
\end{equation}
\section{Proposed Detectors}\label{Spropos}
We use the LRT to derive test statistics for the BHT in (\ref{eq5}). To proceed, based on the i.i.d. property and CSCWG distribution, the log-likelihood function (LLF) of the received samples in the FC during $N$ samples can be written as
\begin{equation}\label{eq6}
\begin{split}
L(\textbf{Y}|H_i) = &-N\log(\pi)-N\log(\det(\textbf{R}_w))\\ 
&-\sum_{n=1}^N (\textbf{y}[n]-\textbf{x}_{i}[n])^H\textbf{R}_w^{-1}(\textbf{y}[n]-\textbf{x}_{i}[n]).
\end{split}
\end{equation}
 The $H$ superscript denotes the hermitian operation. Using the trace of a matrix and its properties, we can write
\begin{equation}\label{eq7}
\begin{split}
L(\textbf{Y}|H_i) = &-N\log(\pi)-N\log(\det(\textbf{R}_w))\\
&-\mathrm{tr}(\textbf{R}_w^{-1}(\textbf{Y}-\textbf{X}_{i})(\textbf{Y}-\textbf{X}_{i})^H).
\end{split}
\end{equation}
Using the LLF and parameterizing it with the unknown parameters in each hypothesis, denoted by $\boldsymbol{\theta}_i$, the LRT becomes
\begin{equation}\label{eq8}
T_{LRT}(\textbf{Y} )= \max_{\boldsymbol{\theta}_1} L(\textbf{Y}|\boldsymbol{\theta}_1,H_1)-\max_{\boldsymbol{\theta}_0} L(\textbf{Y}|\boldsymbol{\theta}_0,H_0)\begin{array}{c}
    H_1  \\
     \gtrless \\
     H_0
\end{array}  \eta,
\end{equation}
where the maximum likelihood estimation (MLE) of unknown parameters is used to derive the LRT and $\eta$ is a detection threshold. Integrating the MLE of unknown parameters into the LRT is also referred to as generalized LRT (GLRT). It is important to note that both the GLRT and MLE can asymptotically achieve the optimal performance \cite{kay1993fundamentalsDetectV2, kay-estimation}, which is the main reason we employ these methods in this work to develop the tests.\par
In the following, we solve the BHT with various assumptions on the availability of the statistical information of noise (SIN) and signal structure in the receiver side.
\subsection{Case1: Available SIN  in SNs}\label{M1Case1}
If SIN including noise variance in all SNs is available in the FC in the BHT in  (\ref{eq5}),  the unknown parameter space under each hypothesis is $\boldsymbol{\theta}_i=\{\textbf{X}_i\}$ for $i=0,1$. Additionally, by considering the low-rank structure for matrices $\textbf{X}_i$, the maximum of the LLF  under $H_i$ is  $\max_{
     \textbf{X}_i }L(\textbf{Y}|\textbf{X}_i,H_i)$ subject to constraint  $C_{M,i}: \mathrm{rank}(\textbf{X}_i)=M+i$. We also assume that $\textbf{R}_w=\sigma^2_w\textbf{I}$, and then we show that the derived LRT is able to handle arbitrary structures for $\textbf{R}_w$ with a straightforward transformation. Now, using (\ref{eq7}) and ignoring non-relevant terms to $\textbf{X}_0$, the equivalent problem is 
\begin{equation}\label{eq9}
 L_0= \min_{\tiny\begin{array}{c}
     \textbf{X}_0 \\
      s.t. C_{M,0}
 \end{array}}\frac{1}{\sigma^2_w}\mathrm{tr}(\textbf{Z}_0 \textbf{Z}_0^H).
\end{equation}
If $\textbf{Z}_0=\textbf{Y}-\textbf{X}_0$, we can use the properties of the  trace to write $\mathrm{tr}(\textbf{Z}_0\textbf{Z}_0^H)=\sum_{k=1}^K\lambda_k^2(\textbf{Z}_0)$, where $\lambda_k(\textbf{Z}_0)$ are the ordered singular values of matrix $\textbf{Z}_0$.
By employing singular value decomposition (SVD), we can write $\textbf{Y}=\sum_{k=1}^K\lambda_k(\textbf{Y})\textbf{u}_k\textbf{v}_k^H$, where $\textbf{u}_k$ and $\textbf{v}_k$ are the corresponding eigenvectors and $\lambda_1(\textbf{Y})\geq,...\geq\lambda_K(\textbf{Y})$. It is straightforward to show that the optimum estimated matrix for $\textbf{X}_0$ with a rank equal to $M$ is $\hat{\textbf{X}}_0=\sum_{k=1}^{M}\lambda_k(\textbf{Y})\textbf{u}_k\textbf{v}_k^H$ since this leads to $\textbf{Z}_0=\textbf{Y}-\hat{\textbf{X}}_0=\sum_{k=M+1}^{K}\lambda_k(\textbf{Y})\textbf{u}_k\textbf{v}_k^H$, and consequently removes the $M$ largest eigenvalue of $\textbf{Y}$ and minimizes the cost function in (\ref{eq9}). Thus, the minimum of (\ref{eq9}) is 
\begin{equation}\label{eq10}
L_0=\min_{\tiny\begin{array}{c}
     \textbf{X}_0 \\
      s.t. C_{M,0}
 \end{array}}\frac{1}{\sigma^2_w}\mathrm{tr}(\textbf{Z}_0 \textbf{Z}_0^H)=\frac{\sum_{k=M+1}^K\lambda_k^2(\textbf{Y})}{\sigma^2_w}.
\end{equation}

Similarly, for $H_1$ with the constraint $\mathrm{rank}(\textbf{X}_1)=M+1$, we can conclude that the MLE for this matrix  is $\hat{\textbf{X}}_1=\sum_{k=1}^{M+1}\lambda_k(\textbf{Y})\textbf{u}_k\textbf{v}_k^H$, resulting  in 
\begin{equation}\label{eq11}
 L_1=\min_{\tiny\begin{array}{c}
     \textbf{X}_1 \\
      s.t. C_{M,1}
 \end{array}}\frac{1}{\sigma^2_w}\mathrm{tr}(\textbf{Z}_1 \textbf{Z}_1^H)=\frac{\sum_{k=M+2}^K\lambda_k^2(\textbf{Y})}{\sigma^2_w},
\end{equation}
where $\textbf{Z}_1=\textbf{Y}-\textbf{X}_{1}$. Plugging (\ref{eq10}), (\ref{eq11}), and (\ref{eq7}) into (\ref{eq8}) and performing a few mathematical manipulations, we get the LRT as
\begin{equation}\label{eq12}
T_{LRT}(\textbf{Y} )= L_0 -L_1\begin{array}{c}
    H_1  \\
     \gtrless \\
     H_0
\end{array}  \eta,
\end{equation}
and consequently, the LRT is
\begin{equation}\label{eq13_ksv}
\begin{split}
&T_{LRT}(\textbf{Y} )=\\ &\frac{\sum_{k=M+1}^K\lambda_k^2(\textbf{Y})}{\sigma^2_w}-\frac{\sum_{k=M+2}^K\lambda_k^2(\textbf{Y})}{\sigma^2_w}=\frac{\lambda_{M+1}^2(\textbf{Y})}{\sigma^2_w}\begin{array}{c}
    H_1  \\
     \gtrless \\
     H_0
\end{array}  \eta.
\end{split}
\end{equation}
The resulting test is based on the $k^{th}$-largest singular value (KSV) where $k=M+1$. To use this detector, it is necessary to have $K>M$. It is interesting to note that if only one TN is active, i.e., $M=1$,  the KSV test becomes
\begin{equation}\label{eq13_new}
T_{LRT}(\textbf{Y} )=\frac{\lambda_{2}^2(\textbf{Y})}{\sigma^2_w}\begin{array}{c}
    H_1  \\
     \gtrless \\
     H_0
\end{array}  \eta.
\end{equation}
We refer to this detector as the second-largest singular value (SSV), where it should be compared with a detection threshold $\eta$ to decide on the presence of a jamming signal. Similarly, to employ the SSV detector, it is necessary to have $K>1$  SNs in the network.\par
Although we assume that $\textbf{R}_w=\sigma^2_w\textbf{I}$, if SIN is available, the proposed SSV jamming detector can also be utilized for scenarios where the noise covariance is arbitrary and the noise variance differs at the SNs, i.e., $\textbf{R}_w\neq \sigma^2_w\textbf{I}$.
In FC,  a transformation on the received signal matrix as a preprocessing can be performed such that  $\textbf{Y}'=\textbf{LX}_i+\textbf{LW}$, where $\textbf{R}_w^{-1}=\textbf{L}^T\textbf{L}$ to equalize the noise variance in each node and to force the columns of $\textbf{LW}$ to follow $\mathcal{CN}(\textbf{0},\textbf{I})$. Since the transformation matrix of $\textbf{L}$ is a full rank matrix, the rank of the transformed matrices $\textbf{LX}_i$ does not change, and similarly, the SSV jamming detector is the LRT for the transformed signal of $\textbf{Y}'$. Since with this transformation, we have $\sigma^2_2=1$, the SSV formula in (\ref{eq13_new}) becomes $\lambda_2^2(\textbf{Y}')$ for the transformed received matrix.

\subsection{Case2: Unavailable SIN with equal $\sigma^2_{w}$ in SNs  }\label{M1case2}
Here, we assume that the noise variance of the collected samples in the FC are identical, $\textbf{R}_w=\sigma^2_w\textbf{I}$, while $\sigma_w^2$ is unknown by the FC. Therefore, the current problem is more complex than the previous case since its unknown parameter space contains the noise variance as an extra unknown parameter; $\boldsymbol{\theta}_i=\{\textbf{X}_i, \sigma^2_w\}$. We also impose the following low-rank structure as a constraint, denoted by $C_{M,i}$, for finding the MLE
\begin{equation}\label{newEq13}
   C_{M,i}: \mathrm{rank}(\textbf{X}_i)=M+i.
\end{equation}
Hence, the updated version of the LLF  for the MLE is \begin{equation}\label{eq13}
\begin{split}
 \max_{\tiny\begin{array}{c}
     \sigma^2_w, \textbf{X}_i \\
      s.t. \ C_{M,i}
     \end{array}}L&(\textbf{Y}|\textbf{X}_i, \sigma^2_w,H_i) = \\
     \max_{\tiny\begin{array}{c}
     \sigma^2_w, \textbf{X}_i \\
      s.t. \ C_{M,i}
     \end{array}}\{&-N\log(\pi)-NK\log(\sigma_w^2)\\
     &-\frac{1}{\sigma^2_w}\mathrm{tr}((\textbf{Y}-\textbf{X}_{i})(\textbf{Y}-\textbf{X}_{i})^H)\}.
     \end{split}
\end{equation}
Employing the same method as in the available SIN case results in an identical maximum MLE for $\textbf{X}_i$.
Therefore, by plugging $\hat{\textbf{X}}_i$, obtained in Section \ref{M1Case1},  into the LLF and substituting  (\ref{eq10})  and  (\ref{eq11})  into (\ref{eq13}), we can write
\begin{equation}\label{eq14}
\begin{split}
  \max_{
     \sigma^2_w} &\ L(\textbf{Y}|\hat{\textbf{X}}_i, \sigma^2_w,H_i) = \\
    &\max_{
     \sigma^2_w}\{-N\log(\pi)-NK\log(\sigma_w^2)-\frac{\sum_{k=M+i+1}^K\lambda_k^2(\textbf{Y})}{\sigma^2_w}\}.
     \end{split}
\end{equation}
To proceed, we need  to determine the MLE of  $\sigma^2_w$ under each hypothesis by solving  $\partial L(\textbf{Y}|\hat{\textbf{X}}_i, \sigma^2_w,H_i)/\partial \sigma^2_w=0$. It is straightforward to solve this equation and find the MLE under each $H_i$ as $\hat{\sigma}^2_{w,i}=\sum_{k=M+i+1}^K\lambda_k^2(\textbf{Y})/NK$. Consequently, we can find the LLF for each hypothesis as 
\begin{equation}\label{eq15}
L(\textbf{Y}|\hat{\textbf{X}}_i, \hat{\sigma}^2_{w,i},H_i)=-N\log(\pi)-NK\log(\hat{\sigma}^2_{w,i})-NK.    
\end{equation} 
Now, by plugging (\ref{eq15}) for each hypothesis into (\ref{eq8}), the LRT is $T_{LRT}(\textbf{Y})=NK\log(\hat{\sigma}^2_{w,0}/\hat{\sigma}^2_{w,1})$, and with simple manipulations, we have 
\begin{equation}\label{eq16}
T_{LRT}(\textbf{Y})= \frac{\sum_{k=M+1}^K\lambda_k^2(\textbf{Y})}{\sum_{k=M+2}^K\lambda_k^2(\textbf{Y})} \begin{array}{c}
    H_1  \\
     \gtrless \\
     H_0
\end{array}  \eta,
\end{equation} 
or equivalently 
\begin{equation}\label{eq16dum}
T_{LRT}(\textbf{Y})= \frac{\lambda_{M+1}^2(\textbf{Y})}{\sum_{k=M+2}^K\lambda_k^2(\textbf{Y})} \begin{array}{c}
    H_1  \\
     \gtrless \\
     H_0
\end{array}  \eta.
\end{equation}
We refer to this test as the generalized ratio of singular values (GRSVs). To perform detection with this test, the condition $K>M+1$ needs to be satisfied.\par
For the specific case where only one TN is active $M=1$, the GRSV reduces to the following test
\begin{equation}\label{eq22_new}
T_{LRT}(\textbf{Y})= \frac{\lambda_{2}^2(\textbf{Y})}{\sum_{k=3}^K\lambda_k^2(\textbf{Y})} \begin{array}{c}
    H_1  \\
     \gtrless \\
     H_0
\end{array}  \eta.
\end{equation}
Since this test statistic is based on the ratio of singular values (RSVs),  we refer to it as the RSV test.  Importantly,  given the fact that these tests are derived based on GLRT, they require no prior information on the variance of the noise. This leads to a desirable property called the constant false alarm rate (CFAR) property \cite{kay1993fundamentalsDetectV2}. In Section VI, we also show the CFAR property of this detector through simulations.
\subsection{Case3: Detection Based on Energy of Signal}
In the literature, numerous studies make use of the energy of the signal or noise to detect the presence of a transmitted signal from a jammer or a primary user. Instead of assuming $\textbf{x}_i[n]$ as unknown deterministic parameters \cite{Cooperative_WED},  here, we consider one TN ($M=1$), and we assume that the channel gains are not fixed during each sensing time, i.e., $\textbf{h}_{s,1}[n] \sim \mathcal{CN}(\textbf{0},\sigma^2_{h_{s}}\textbf{I})$ and $\textbf{h}_j[n] \sim \mathcal{CN}(\textbf{0},\sigma^2_{h_{j}}\textbf{I})$. Consequently, we can write the BHT problem for jamming detection using the i.i.d. property of the noise samples as follows
\begin{equation}\label{eq19}
\Big{\{}\begin{array}{c}
     H_0 : \textbf{y}[n] \sim \mathcal{CN}(\textbf{0},P_s\sigma^2_{h_{s}}\textbf{I}+ \sigma_w^2\textbf{I}) \ \ \ \ \ \ \ \ \ \  \ \\
     H_1 : \textbf{y}[n] \sim \mathcal{CN}(\textbf{0},P_s\sigma^2_{h_{s}}\textbf{I}+P_j\sigma^2_{h_{j}}\textbf{I}+ \sigma_w^2\textbf{I}) 
\end{array},
\end{equation}
where  $n = 1, ..., N$, and $P_s$ and $P_j$ denote the power of the transmitted signal by TN and JN, respectively. We also assumed $\textbf{R}_w=\sigma_w^2\textbf{I}$. If all the statistical parameters of the signals, such as the signal power of the JN and the channel gains between the JN and the SNs, are known in the FC, the Nayman-Person (NP) theorem for the BHT in (\ref{eq19}) results in the well-known test statistic, called energy detector (ED) \cite{kay1993fundamentalsDetectV2}. However, we assume that $P_j$ and $\sigma^2_{h_{j}}$ are unknown in the FC, and we attempt to find the locally most powerful (LMP) test for the problem  (\ref{eq19}). To proceed, we define the test parameter $\sigma^2_y$, and we rewrite (\ref{eq19}) as follows
\begin{equation}\label{eq20}
\Big{\{}\begin{array}{c}
     H_0 :\sigma^2_y = \sigma^2_{H_0} \\
     H_1 : \sigma^2_y > \sigma^2_{H_0}
\end{array},
\end{equation}
where $\sigma^2_{H_0}=P_s\sigma^2_{h_{s}}+ \sigma_w^2$ is known at the receiver.
To find the LMP test, we use \cite{kay1993fundamentalsDetectV2}
\begin{equation}\label{eq21}
    T_{LMP}(\textbf{Y})= \frac{ \frac{\partial L(\textbf{Y}| \sigma^2_y)}{ \partial \sigma^2_y}|_{\sigma^2_y = \sigma^2_{H_0}}}{ \sqrt {I(\sigma^2_{H_0}) }},
\end{equation}
where $I(\sigma^2_{H_0})$ is the Fischer information. In Appendix A, we show that the LMP test for this problem is equivalent to ED 
\begin{equation}\label{eq22}
    T_{LMP}(\textbf{Y})= \sum_{n=1}^N||\textbf{y}[n]||^2_2,
\end{equation}
where $||.||_2$ is the 2-norm of the vector. We take the ED into consideration in the rest of this work for comparisons since it is  used for jamming detection problems in cooperative \cite{Arcangeloni,Cooperative_WED} and multicarrier \cite{kurt} scenarios. ED requires prior information on the transmitted signal from TN in addition to the noise information. This is why several works proposed energy-based detectors for jamming and spoofing attacks detection during pilot transmission, where the friendly transmitted signal is known to the receiver \cite{JAD, VAR,Enenrgy_ratio_detection}.

\section{Multiple Number of JNs}\label{S_MJN}
Here, we regard scenarios in which more than one JN is active to examine the performance of detectors not only under multiple TNs but also with more than a single active JN. For simplicity, let $M=1$ and $\sigma^2_w=1$, and we only consider SSV  with the formula  $T(\textbf{Y})=\lambda^2_2(\textbf{Y})$ since the same method can be extended to other proposed detectors. \par
We rewrite the new detection problem in  (\ref{eq4})  with two JNs and three hypotheses as 
\begin{equation}\label{eq1mjn}
 \left\{ \begin{array}{c} 
     H_0 : \textbf{Y} =\textbf{X}_0 + \textbf{W} \\
     H_1 : \textbf{Y} = \textbf{X}_1+\textbf{W} \\
     H_2 : \textbf{Y} = \textbf{X}_2+\textbf{W}
\end{array},\right.
\end{equation}
where $\textbf{X}_0=\textbf{h}_{s}\textbf{s}^T$, 
$\textbf{X}_1=\textbf{h}_{s}\textbf{s}^T+\textbf{h}_{j,1}\textbf{j}_1^T$, $\textbf{X}_2=\textbf{h}_{s}\textbf{s}^T+\textbf{h}_{j,1}\textbf{j}_1^T+\textbf{h}_{j,2}\textbf{j}_2^T$. Here, $\textbf{h}_{j,1}$, $\textbf{h}_{j,2}$, $\textbf{j}_1$, and $\textbf{j}_2$ are the channel vectors for JN1 and JN2 and the transmitted vectors  by JN1 and JN2, respectively. Importantly,  we are only interested in detecting the jamming attack and not finding the number of JNs. Therefore, we try to realize whether $H_0$ (no jamming attack) or $\{H_1, H_2\}$ is confirmed without considering the number of JNs. \par
If  we regard the large signal-to-noise ratio (SNR) regime,  where we can ignore the noise in equations for mathematical simplicity, we can write  the received signal under a jamming attack by two JNs as
\begin{equation}\label{eq2mjn}
\textbf{Y}\approx \textbf{X}_2=\textbf{h}_{s}\textbf{s}^T+\textbf{h}_{j,1}\textbf{j}_1^T+\textbf{h}_{j,2}\textbf{j}_2^T.
\end{equation}
For convenience, we assume that all channel vectors between nodes are orthogonal to each other, and signal vectors from the TN and the JNs are also orthogonal. The orthogonality assumption simplifies the relation between the norms of each signal with its singular values. However, in the simulations, we provide the result of the general scenario without the orthogonality assumption. As a result, by orthogonality assumption, each vector can be an eigenvector for matrix $\textbf{X}_2$, and the SVD will be
\begin{equation}\label{eq3mjn}
\begin{split}    
&\textbf{Y}\approx \textbf{X}_2=\\
&\|\textbf{h}_{s}\|_2\|\textbf{s}\|_2\bar{\textbf{h}}_{s}\bar{\textbf{s}}^T
+\|\textbf{h}_{j,1}\|_2\|\textbf{j}_1\|_2\bar{\textbf{h}}_{j,1}\bar{\textbf{j}}_1^T
+\|\textbf{h}_{j,2}\|_2\|\textbf{j}_2\|_2\bar{\textbf{h}}_{j,2}\bar{\textbf{j}}_2^T.
\end{split}
\end{equation}
where the bar on each vector shows that the vector is normalized.  Consequently, we can introduce  $\lambda_s(\textbf{X}_2)=\|\textbf{h}_{s}\|_2\|\textbf{s}\|_2$,  $\lambda_{j_1}(\textbf{X}_2)=\|\textbf{h}_{j,1}\|_2\|\textbf{j}_1\|_2$, and $\lambda_{j_2}(\textbf{X}_{2})=\|\textbf{h}_{j,2}\|_2\|\textbf{j}_2\|_2$
as three singular values while we are unaware of their order. If we assume that $\lambda_s(\textbf{X}_2)$ is the greatest one, then we can simply write the probability of detection for the  SSV under the orthogonality assumption as 
\begin{equation}\label{eq4mjn}
P^{o}_{d|H_2}=\mathrm{Pr}[\lambda^2_2(\textbf{Y})\approx \max\{\lambda^2_{j_1}(\textbf{X}_2), \lambda^2_{j_2}(\textbf{X}_2)\}>\eta|H_2]
\end{equation}
where superscript ``\textit{o}'' denotes that the probability is computed under the orthogonality assumption. If we denote the probability of detection under $H_1$ and the orthogonality assumption as $P^{o}_{d|H_1}=\mathrm{Pr}[\lambda^2_2(\textbf{Y})\approx \lambda^2_{j_1}(\textbf{X}_2)>\eta|H_1]$, then we can write equation (\ref{eq4mjn})  as 
\begin{equation}\label{eq5mjn}
\begin{split}
&P^{o}_{d|H_2}=\\
&P^{o}_{d|H_1}+\mathrm{Pr}[\lambda^2_{j_2}(\textbf{X}_2)>\eta|H_2]-P^{o}_{d|H_1}\mathrm{Pr}[\lambda^2_{j_2}(\textbf{X}_2)>\eta|H_2].
\end{split}
\end{equation}
If $\|\textbf{h}_{j,1}\|_2\|\textbf{j}_1\|_2=\|\textbf{h}_{j,2}\|_2\|\textbf{j}_2\|_2$, equation (\ref{eq5mjn}) reduces to 
\begin{equation}\label{eq6mjn}
\begin{split}
P^{o}_{d|H_2}=2P^{o}_{d|H_1}-(P^{o}_{d|H_1})^2.
\end{split}
\end{equation}
This simply shows the improvement in the probability of detection of the SSV by the presence of the second JN and the ability of this detector to handle the detection of attack by multiple JNs.  Interestingly, if $\lambda_{j_2} \rightarrow 0$, then we will be under $H_1$, and the probability of detection becomes equal to $P^{o}_{d|H_1}$. Additionally, if we assume any of the signals from the JNs are the largest, similar results can be observed. In the following, we show that if we discard the orthogonality assumption and instead assume that the JNs' channel vectors and symbol vectors are correlated, this also enhances the detection performance of the SSV.

\section{Discussion on False Alarms by SSV and KSV and Detection Threshold Adjustment}\label{Sfalse}
% \begin{figure}[!tb]
%     \centering
%     \includegraphics[width =3.5in, height=2.45in]{Pfa_eta_Analytic_MC_gs=0dB-eps-converted-to.pdf}
%     \caption{Probability of false alarm of the SSV detector versus $\eta$ based on Monte Carlo (MC) simulation and analytical (A) formula with $M=1$, $K=3$, and $\gamma_s=0$ dB }
%     \label{fig01}
% \end{figure}
\subsection{Analytical Probability of  False Alarm}
Here, we concentrate on the proposed SSV and KSV detectors and their probability of false alarm from the analytical side, which is obtained as $P_{fa}=\mathrm{Pr}\{T(\textbf{Y})>\eta|H_0\}$.\par
The distribution of eigenvalues of matrices following a Wishart distribution has been extensively studied in the wireless communication literature \cite{Zanella2003,Zanella2009}. Prior to using this distribution, it is noteworthy that by assuming $\sigma_w^2=1$, the SSV is equivalent to $\lambda_2(\textbf{Y}\textbf{Y}^H)$, which is the second largest eigenvalue of the matrix $\textbf{Y}\textbf{Y}^H$. Under $H_0$, in contrast to spectrum sensing problems, the desired signal still exists which brings more complexity to the analysis of the distribution of the SSV detector. Moreover, it should be mentioned that although assuming fixed channel gains during a sensing time with $N$ samples provides a low-rank structure for $\textbf{X}_0$, it might also cause a correlation between the columns of $\textbf{X}_0$. To capture the correlation between the columns of $\textbf{Y}$, we write its  covariance matrix as 
\begin{equation}\label{eq31n}
\begin{split}
    \textbf{R}_{\textbf{Y}}=
    &\frac{1}{N}\sum_{n=1}^N \mathrm{E}[\textbf{y}[n]\textbf{y}[n]^H] =\\ &\frac{1}{N}\sum_{n=1}^N \mathrm{E}[\textbf{x}_0[n]\textbf{x}_0[n]^H] +\frac{1}{N}\sum_{n=1}^N\mathrm{E}[\textbf{w}[n]\textbf{w}[n]^H],
\end{split}
\end{equation}
where $\mathrm{E}[\cdot]$ is the expected value, and we assume $\frac{1}{N}\sum_{n=1}^N\mathrm{E}[\textbf{x}_0[n]\textbf{w}[n]^H]=0$. Using $\textbf{x}_0[n]=\sum_{m=1}^M\textbf{h}_{s,m}s_m[n]$ and assuming that  $\textbf{h}_{s,m}$ is a deterministic parameter and $s_m$ is a random variable with $s_m\sim \mathcal{CN}(0,P_s)$, equation (\ref{eq31n}) can be rewritten as
\begin{equation}\label{eq32n}
\textbf{R}_{\textbf{Y}}=\sum_{m=1}^M\sum_{l=1}^M\textbf{h}_{s,m}\textbf{h}^H_{s,l}\frac{1}{N}\sum_{n=1}^N \mathrm{E}[s_m[n]s_l^*[n]]+\sigma_w^2\textbf{I}.
\end{equation}
Assuming that $\frac{1}{N}\sum_{n=1}^N \mathrm{E}[s_m[n]s_l^*[n]]=P_s\delta [m-l]$, where $\delta[\cdot]$ is the discrete Delta function, we can further simplify (\ref{eq32n}) as
\begin{equation}\label{eq33n}
\textbf{R}_{\textbf{Y}}=P_s\sum_{m=1}^M\textbf{h}_{s,m}\textbf{h}^H_{s,m}+\sigma_w^2\textbf{I}= \textbf{R}_{\textbf{X}_0}+\sigma_w^2\textbf{I},
\end{equation}
where $\mathrm{rank}(\textbf{R}_{\textbf{X}_0})=M$.\par
Since the distribution of $\textbf{y}[n]$ is a complex normal distribution, the matrix $\textbf{YY}^H$ follows a complex Wishart distribution \cite{Zanella2003}. Therefore, we employ the distribution of eigenvalues of the Wishart matrix to find the probability of false alarm for the SSV detector $P_{fa}=\mathrm{Pr}\{\lambda_2(\textbf{Y}\textbf{Y}^H)>\eta|H_0\}$. As $\textbf{R}_{\textbf{Y}}$ does not have a diagonal format,  illustrating the existing correlation of columns, and we assume  $\mathrm{E}\{\textbf{y}[n]\}=\textbf{0}$, the joint probability density function (PDF) of the eigenvalues of  the correlated central Wishart matrix is given by \cite{Zanella2003,Zanella2009}
\begin{equation}\label{eq25}
    f_{\boldsymbol{\lambda}}(\lambda_1, ... , \lambda_k)=c_0|\textbf{E}(\boldsymbol{\lambda},\boldsymbol{\zeta})||\textbf{V}(\boldsymbol{\lambda})|\prod_{k=1}^{K}\lambda_k^{N-K},
\end{equation}
which is helpful for formulating the $P_{fa}$ of the SSV and KSV detectors. In equation (\ref{eq25}), $c_0$ is a normalizing constant and $\textbf{V}(\boldsymbol{\lambda})$ is a Vandermode matrix  whose $(i,j)^{th}$ element is $[\textbf{V}(\boldsymbol{\lambda})]_{i,j}=\lambda_j^{i-1}$. The determinant of a matrix is shown by the operator  $|.|$. In addition, $  \textbf{E}(\boldsymbol{\lambda},\boldsymbol{\zeta})$ is a matrix, where $[\textbf{E}(\boldsymbol{\lambda},\boldsymbol{\zeta})]_{i,j} = e^{-\lambda_j/\zeta_i}$, and $\zeta_i$ are the ordered eigenvalues ($\zeta_1> ...>\zeta_K$) of  $\textbf{R}_{\textbf{Y}}$.  Now, using equation (\ref{eq25}), we continue to formulate the probability of false alarm of the SSV and KSV detectors.
\subsubsection{Example with three SNs and one TN}
 %%%%%%%%%%%%%%%%%%%%%%%%%%%%%%%
% \begin{figure}[!tb]
%     \centering
%     \includegraphics[width =3.5in, height=2.45in]{Pfa_KSV_analytic_gammas0db_k3-eps-converted-to.pdf}
%     \caption{Probability of false alarm of the KSV detector versus $\eta$ based on Monte Carlo (MC) simulation and analytical (A) formula with $M=2$, $K=3$, and $\gamma_s=0$ dB }
%     \label{fig01n}
% \end{figure}
 %%%%%%%%
Here, without loss of generality and for the sake of simplicity, we consider the scenario with $K=3$. Thus, the $P_{fa}$ of the SSV detector can be analytically calculated  as
\begin{equation}\label{eq26}
\begin{split}
    P_{fa}&= 1-\mathrm{Pr}\{\lambda_2(\textbf{Y}\textbf{Y}^H)<\eta|H_0\}\\ &=1-
    \int_0^\eta \int_{\lambda_2}^{\infty}\int_{0}^{\lambda_2}f_{\boldsymbol{\lambda}}(\lambda_1, ... , \lambda_3)d\lambda_3 d\lambda_1 d\lambda_2.
\end{split}
\end{equation}
To compute the probability using this formula, it is important to note that the eigenvalues, $\lambda_i$, should differ to guarantee that $|\textbf{E}(\boldsymbol{\lambda},\boldsymbol{\zeta})| \neq 0$. As a result, we use the ordered eigenvalues of  $\textbf{R}_{\textbf{Y}}$, which are included in the vector $\boldsymbol{\zeta}=[\zeta_1, \zeta_2, \zeta_3]= [\xi_1+\sigma_w^2,\epsilon+\sigma_w^2,\sigma_w^2 ]$ where $\mathrm{rank}(\textbf{R}_{\textbf{X}_0})=M=1$, and $\xi_1$ is the only eigenvalue of  $\textbf{R}_{\textbf{X}_0}$. Since $\textbf{R}_{\textbf{Y}}$ in (\ref{eq33n}) comprises $\textbf{R}_{\textbf{X}_0}$ with rank one and the diagonal matrix $\sigma_w^2\textbf{I}$, in the vector $\boldsymbol{\zeta}$, a small amount, $\epsilon$, is added to $\zeta_2$ in order to have $\zeta_2>\zeta_3$. Of note, in order to use equation (\ref{eq26}) to adjust $\eta$ for the SSV detector, we need to know or estimate $\xi_1$ and $\sigma_w^2$ at the FC. 
\subsubsection{Example with three SNs and two TNs}
In the case with $M=2$, if $\sigma_w^2=1$, the KSV detector using  equation (\ref{eq13_ksv}) can be found based on $\lambda^{2}_3(\textbf{Y})$ or equivalently  based on $\lambda_3(\textbf{YY}^H)$.  Therefore, to analyze the $P_{fa}$ of the KSV detector, we can use equation (\ref{eq25}) to find the PDF of  $\lambda_3(\textbf{YY}^H)$ as follows
\begin{equation}\label{eq36n}
\begin{split}
    P_{fa}&= 1-\mathrm{Pr}\{\lambda_3(\textbf{Y}\textbf{Y}^H)<\eta|H_0\}\\ &=1-
    \int_0^\eta \int_{\lambda_3}^{\infty}\int_{\lambda_2}^{\infty}f_{\boldsymbol{\lambda}}(\lambda_1, ... , \lambda_3)d\lambda_1 d\lambda_2d\lambda_3,
\end{split}
\end{equation}
where $\mathrm{rank}(\textbf{R}_{\textbf{X}_0})=M=2$, and the ordered eigenvalues of  $\textbf{R}_{\textbf{Y}}$  are $\boldsymbol{\zeta}=[\zeta_1, \zeta_2, \zeta_3]= [\xi_1+\sigma_w^2,\xi_2+\sigma_w^2,\sigma_w^2 ]$ where $\xi_i$ denotes the ordered  eigenvalues of  $\textbf{R}_{\textbf{X}_0}$.\par
%-----------------
%----------------
\subsection{Detection Threshold Adjustment}
The detection threshold $\eta$ should be selected so that the actual false alarm probability $P_{{fa}}$ satisfies $P_{{fa}} \leq p_{{fa}}$, where $p_{{fa}}$ denotes the desired or target probability of false alarm, in order to avoid high $P_{fa}$.  
There are different methods to set the detection threshold depending on the application of the test statistic, such as in radar, spectrum sensing, or jamming detection applications. In what follows, we discuss three methods for threshold selection, each with different required information and limitations.
\subsubsection{Analytical Method}
This method relies on a closed-form or approximated expression relating $P_{{fa}}$ and $\eta$ similar to equations (\ref{eq26}) and (\ref{eq36n}). When such an expression can be derived, it can be used to very accurately adjust $\eta$ without limitations. In addition, in order to employ this method for
 threshold adjustment, it is necessary to have the true values or reliable estimates of all parameters involved in the analytical formula. However, while deriving exact expressions for $P_{{fa}}$ is straightforward for simpler detectors, such as the  ED, which is well-studied in the literature \cite{Axell}, this can be challenging for more complex test statistics, such as the proposed GRSV detector.

\subsubsection{Monte Carlo with Empirical Samples}
% The analytical formula of $P_{fa}$ is beneficial for adjusting the detection threshold $\eta$ and limiting it to $p_{fa}$, 
Besides the analytical formula, another solution is to run  MC simulations over $T_r$ realizations to estimate $P_{{fa}}$. In this case, $T_r$  samples of $\textbf{Y}|H_0$ (the received signals under $H_0$) are required. This set of the received matrix can be collected and sensed experimentally in a jamming-free environment at the implemented receiver side. The collected samples can then be fed to the detector to obtain $T_r$ samples of the detector output under $H_0$, denoted by $T(\textbf{Y})|H_0$. Then, using $P_{{fa}} = T_r^{-1} \sum_{t=1}^{T_r} \mathbb{I}\{T(\textbf{Y}^{(t)}|H_0) > \eta\}$, where $\mathbb{I}\{\cdot\}$ is the indicator function that equals $1$ when the condition in its argument is satisfied, the false alarm probability can be estimated \cite{Gerlach1997,etreme_value}. Here, $ \textbf{Y}^{(t)}|H_0$ denotes the $t^{{th}}$ empirical received matrix at the  FC. This approach requires only a set of jamming-free received signals or samples of $\textbf{Y}|H_0$ at the FC. Therefore, it is easy to use the MC method to plot $P_{{fa}}$ versus  $\eta$ curve and select the desired threshold to satisfy $P_{{fa}} \leq p_{{fa}}$. However, although this approach does not require any prior information, it is not practical when a very small $p_{ fa}$ is required, since accurate threshold setting in that case requires a large number of empirical signal samples.

\subsubsection{Monte Carlo with Synthetic Samples}
In this method, instead of using a set of experimentally received signals under $H_0$, synthetic samples of $\textbf{Y}|H_0$ are generated using a known and accurate statistical model for $\textbf{Y}|H_0$.
 This method inherently needs knowledge of the distribution of $\textbf{Y}|H_0$, as well as information (or reliable estimates) of its associated parameters, including the noise variance, the transmit power of the TN, and the second-order statistics of the channel or CSI. In addition, it also requires pilot signals that can be employed to estimate the CSI and channel gain variance for this purpose. One advantage of this method is that, instead of relying on a limited number of empirical received samples, it can synthetically generate a large number of realizations and more accurately estimate $P_{ fa }$, thereby enabling accurate estimation of $\eta$ even for very small $p_{fa}$. However, this benefit comes at the cost of requiring detailed information about the distribution of $\textbf{Y}|H_0$ and its related parameters.  In practice, obtaining samples of $\textbf{Y}|H_0$ may be challenging. One possible method is to collect these samples through experiments conducted in a controlled environment, such as a laboratory or an isolated setting, that mimics the real-world environment with only active TN.\par  
 
 Furthermore, in a non-stationary environment, with varying statistical parameters such as the channel gain variance $\sigma_{h_s}^2$ or the transmit power by the TN (due to the mobility of the TN or changes in the environment), the MC method requires frequent re-estimation of $\eta$ or those parameters to regenerate new samples of $\textbf{Y}|H_0$. In such cases, pilot signals, CSI estimation techniques, or feedback mechanisms between the FC and TNs can be used to update information about these parameters accordingly. In the next section, we also evaluate robustness of the proposed detectors with respect to variations in key parameters, such as the noise variance and the transmit SNR of the TN. The results reveal that the RSV detector exhibits greater resilience to these changes as compared to the other detectors. \par

Importantly, since we do not have access to empirical signals in our study, we adopt the MC method with synthetic samples to obtain the $P_{{fa}}$  versus $\eta$ curves for all simulated detectors. Yet, it should be emphasized that the MC method using empirical signals is recommended for practical implementations, as this method is simple and requires only a set of received signals under $H_0$ and does not rely on any prior information.
\subsection{Comparison of Analytical and MC Methods for SSV and KSV Detectors}
\begin{figure}[!tb]
    \centering
    \includegraphics[width =3.5in, height=2.45in]{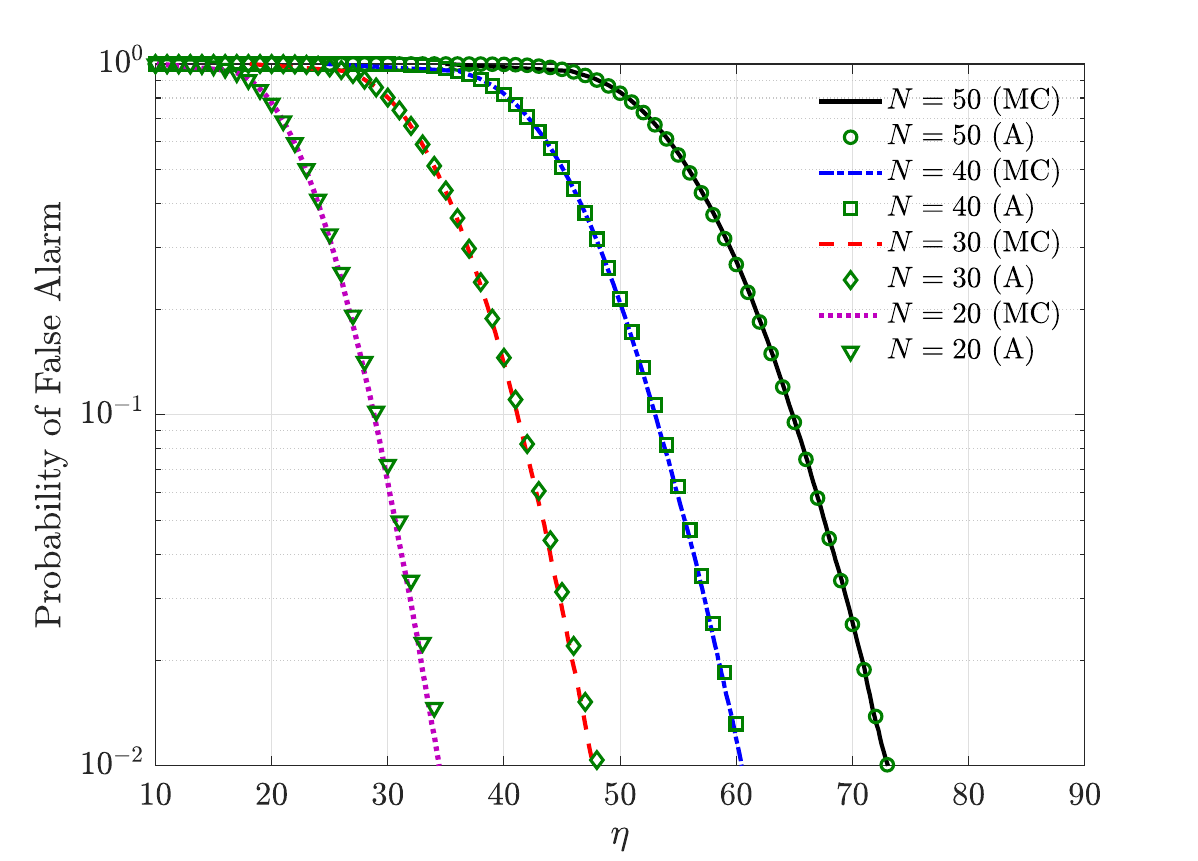}
    \caption{Probability of false alarm of the SSV detector versus $\eta$ based on Monte Carlo (MC) simulation and analytical (A) formula with $M=1$, $K=3$, and $\gamma_s=0$ dB }
    \label{fig01}
\end{figure}
 In the following sections, we compare the analytical curves (e.g., derived from equations \eqref{eq26} and \eqref{eq36n}) with those generated using this MC-based approach. 
% Besides the analytical formula, one can run Monte Carlo (MC) simulations for $T_r$ runs and find $P_{fa}$. In this case the  $T_r$ number of samples of $\textbf{Y}|H_0$ (the received signals under $H_0$) is required this can be sensed experimentally in the envirionment and in practical implemented receiver side and then feeded to the input to have $T_r$ samples of the output of the detector under $H_0$ denoted by $T(\textbf{Y})|H_0$. Then using $P_{fa}=T_r^{-1}\sum_{t=1}^{T_r}\mathbb{I}\{T(\textbf{Y}^{(t)}|H_0)>\eta\}$, where $\mathbb{I}\{.\}$ is the indicator function that is equal to one when the condition in its argument is satisfied. $\textbf{Y}^{(t)}$ denotes the $t^{th}$ empirical received matrix at the FC. This approach only requires a set of received signals $\textbf{Y}$ at the FC under the null hypothesis ($H_0$). Therefore, it is easy to use MC simulation to plot   $P_{fa}$ versus $\eta$  curve and select the desired threshold in order to satisfy $P_{fa}\leq p_{fa}$. In the following comparisons, we use the MC method to adjust the detection threshold for all detectors.\par
In Fig. \ref{fig01}, we plot the  $P_{fa}$ of the SSV detector versus $\eta$ using the analytical formula in equation (\ref{eq26}) and MC simulations. In this simulation, the transmitted signal by  the TN has a transmit SNR, denoted by $\gamma_s = P_s/\sigma^2_w$, which equals $\gamma_s=0$ dB, and the received SNR is denoted by $\gamma_{s,r} = P_s\sigma_{h_s}^2/\sigma^2_w$. As can be seen in the results, the analytical curves are in excellent accordance with the MC simulations. This consistency also exists for various numbers of samples $N$, which confirms the validity of the analytical formula. For larger $\eta$, we observe that the number of false alarms by the SSV detector decreases. However, a very large $\eta$ also results in a lower chance of detection of an active JN by the detector.\par
\begin{figure}[!tb]
    \centering
    \includegraphics[width =3.5in, height=2.45in]{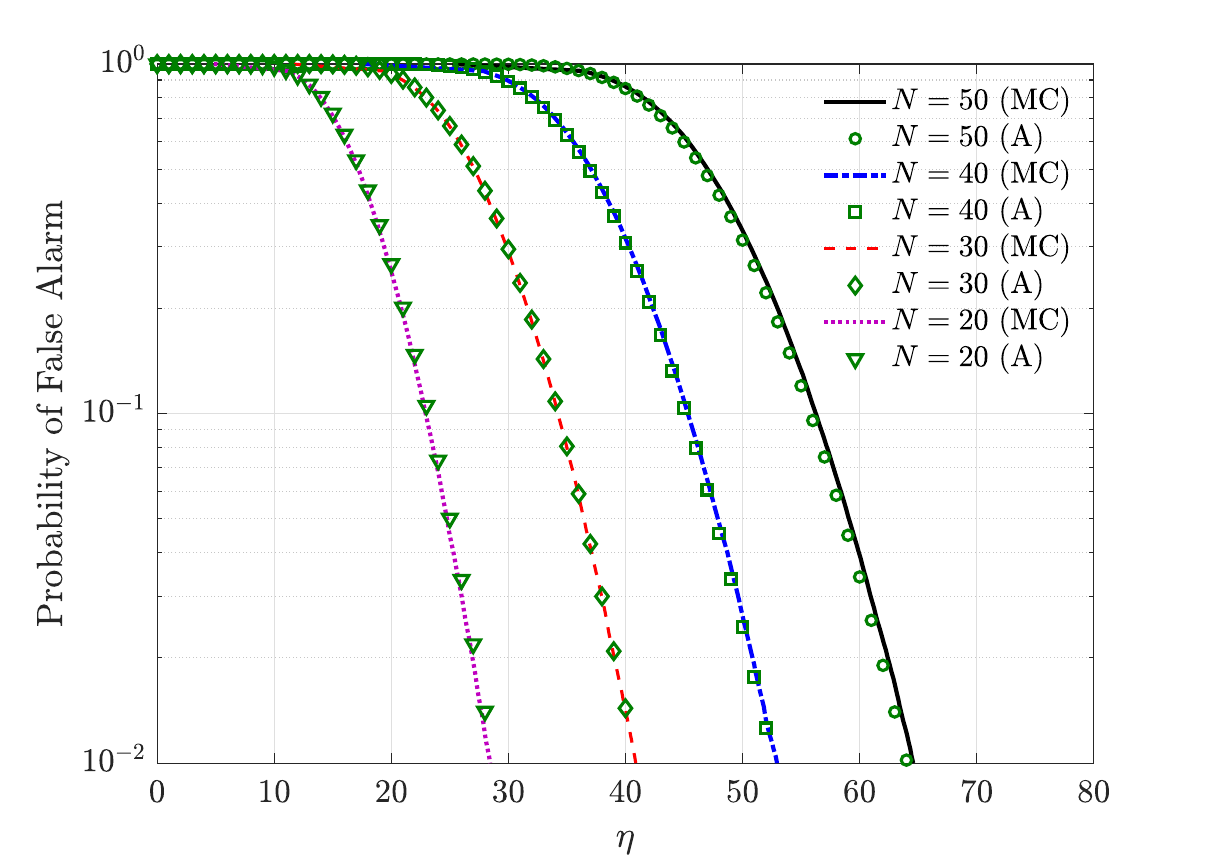}
    \caption{Probability of false alarm of the KSV detector versus $\eta$ based on Monte Carlo (MC) simulation and analytical (A) formula with $M=2$, $K=3$, and $\gamma_s=0$ dB }
    \label{fig01n}
\end{figure}
 Fig. \ref{fig01n} shows analytical curves depicting the probability of false alarm of the KSV detector with $M=2$ and $K=3$ using equation (\ref{eq36n})  versus $\eta$ for different numbers of sensing samples. We set the SNR of both TNs to $\gamma_s=0$ dB. Furthermore,  Fig. \ref{fig01n}  includes false alarm probability curves obtained through MC simulations. These results further confirm the strong agreement between the analytical formulas and MC simulations. Therefore, the analytical method can be reliably used to adjust the detection threshold for the SSV and KSV detectors, thereby eliminating the need for the MC simulations.

\section{Simulations and Numerical Results}\label{Simul}

In this section, we investigate the performance of the proposed detectors, and we compare them with several existing jamming attack detection methods from the literature. Here, two primary metrics are assessed, namely the probability of detection $P_d =\mathrm{Pr}\{T(\textbf{Y})>\eta|H_1\}$ and the probability of false alarm $P_{fa}=\mathrm{Pr}\{T(\textbf{Y})>\eta|H_0\}$. We also define the target or desired probability of false alarm $p_{fa}$, and we adjust $\eta$ for detectors to limit $P_{fa}$ to this upper bound $P_{fa}\leq p_{fa}$ in order to generally restrict the false alarm errors by the detectors. Additional important parameters used in the simulations are the desired received SNR, denoted by $\gamma_{s,r} = P_s\sigma^2_{h_s}/\sigma^2_w$, and the received JN SNR, defined as $\gamma_{j,r} = P_j\sigma^2_{h_j}/\sigma^2_w$. The transmit SNR for the TN is $\gamma_{s} = P_s/\sigma^2_w$ and for the JN is $\gamma_{j} = P_j/\sigma^2_w$. In our simulations, we modeled the channels using random distributions, where new realizations were applied in each iteration. We use QPSK modulation for $s_m[n]$ and also for $j[n]$ to characterize the challenging deceptive jamming attack case. Unless otherwise stated, we set $M=1$, $K=8$, and $N=20$ in the simulated scenarios in this section.\par
%%%%%%%%%%%%%%%%%%%%%%%%%%%%%%%%%%%%%%%%%%%%%%%%%%%%%%%%%%%%%
\begin{figure}[!tb]
    \centering
    \includegraphics[width =3.5in, height=2.5in]{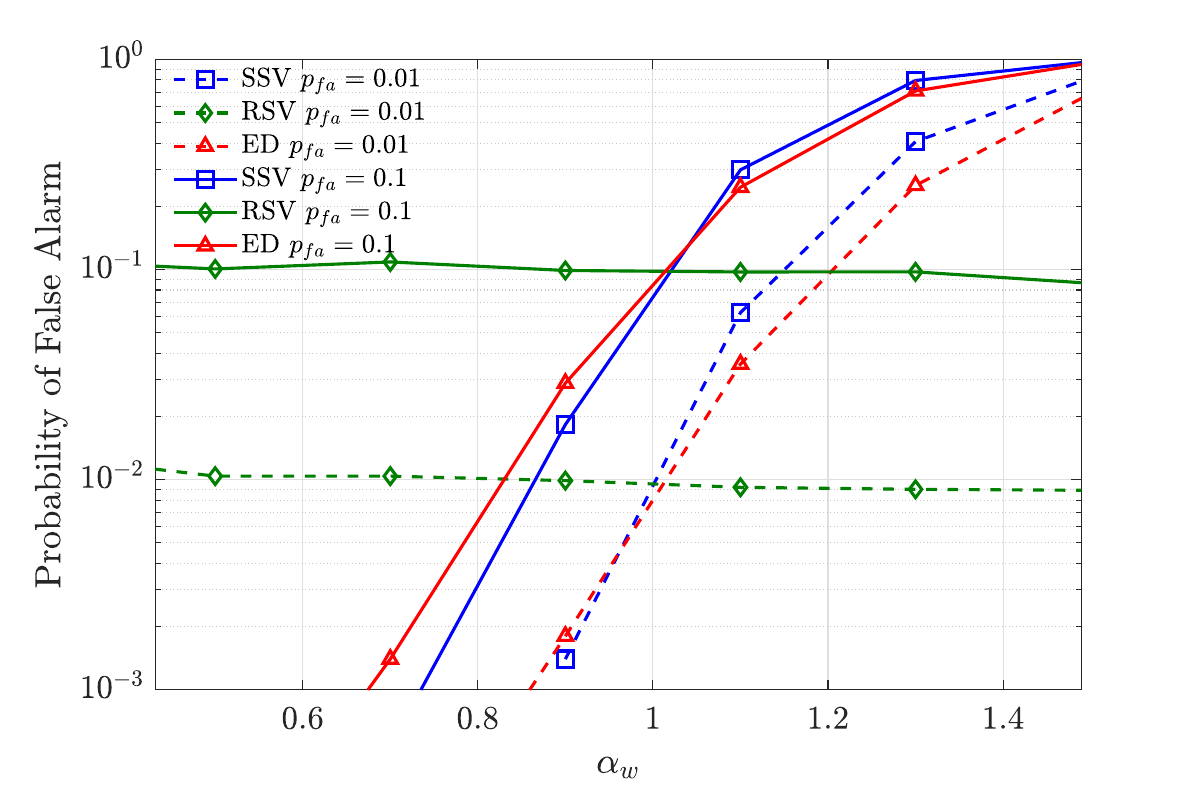}
    \caption{Probability of false alarm versus $\alpha_w$ with $K=8$, $N=20$, and $\gamma_s=5$ dB}
    \label{fig1}
\end{figure}
\begin{figure}[!tb]
    \centering
    \includegraphics[width =3.5in, height=2.5in]{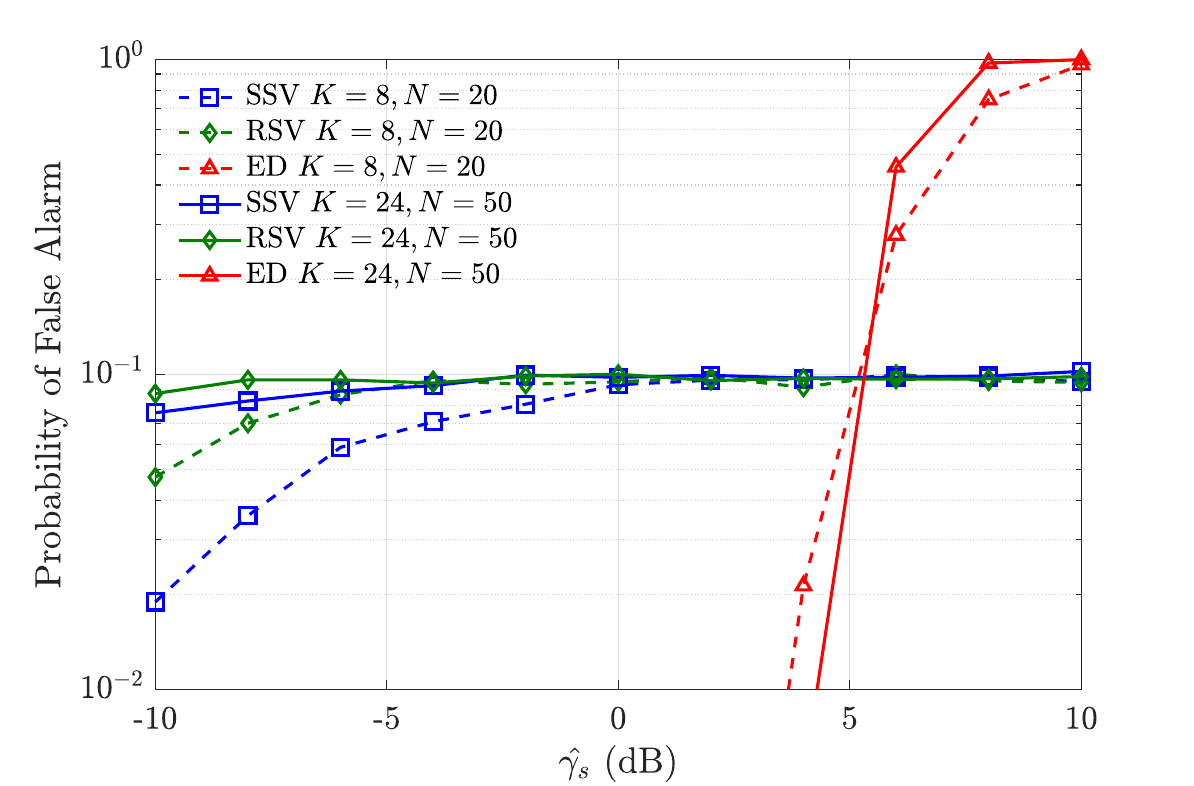}
    \caption{Probability of false alarm versus $\hat{\gamma}_s$ with $p_{fa}=0.1$ when $\gamma_s=5$ dB}
    \label{fig2}
\end{figure}
%%%%%%%%%%%%%%%%%%%%%%%%%%%%%%%%%%%%%%%%%%%%%%%%%%%%%%%%%%%%%%%%%
%----------------------------------------------------
We first simulate the probability of false alarm under $H_0$ for the proposed  SSV and RSV detectors for various parameters,  including $\sigma_w^2$ and $\gamma_s$. 
The ED has also been regarded since it requires information on both these parameters, and as a result,  $P_{fa}$ of the ED is sensitive to these parameters. 
These simulations aim to examine the robustness of the proposed detectors under scenarios where the available information on  $\sigma_w^2$ and $\gamma_s$ at the FC is erroneous and subject to uncertainty. Therefore, we define the considered amount for the noise variance at the FC with $\sigma_w^2$, which is used for the adjustment of the detection threshold. However, the true value of the noise variance, denoted by $\hat{\sigma}_w^2$, is used in the following simulations for the generation of the signals. Consequently, the difference between $\sigma^2_w$ and $\hat{\sigma}_w^2$ captures the error in the known parameters and enables us to examine the robustness of the detectors under uncertainty in the noise variance. Similarly, $\hat{\gamma}_s$ denotes the true value of SNR used for the generation of signals from the TN, while $\gamma_s$ is the considered amount at the FC used for the adjustment of $\eta$.\par
Fig. \ref{fig1} illustrates $P_{fa}$ of the SSV, RSV, and ED detectors as a function of $\alpha_w=\hat{\sigma}_w^2/\sigma^2_w$. % Parameter $\alpha_w$ is the ratio of the noise variance in this simulation ($\hat{\sigma}_w^2$) to the noise variance ($\sigma_w^2$) of the signals under $H_0$ which are used for the adjustment of $\eta$ in MC simulations explained in the previous section. 
 It is clear from this figure that the RSV detector has a constant false alarm rate (CFAR) property with respect to the noise variance while the SSV and ED are susceptible to the noise variance. If $\alpha_w>1$, $P_{fa}$ increases for the SSV and ED detectors, showing higher false alarms. On the other hand, this causes a decrease in $P_{fa}$ and $P_d$ in the case where $\alpha_w<1$, which is also undesirable.\par
 %--------
\begin{figure}[!tb]
    \centering
    \includegraphics[width =\FW]{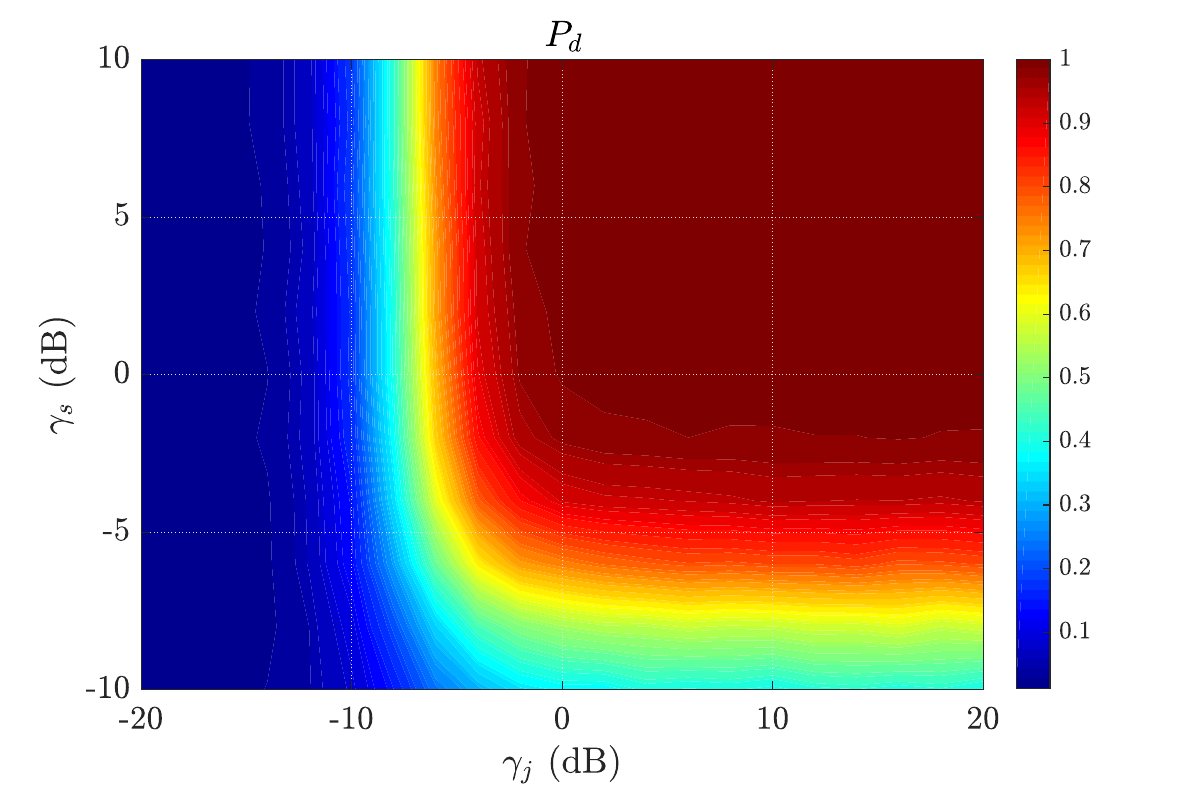}
    \caption{Probability of detection of the SSV detector versus $\gamma_s$ and $\gamma_j$ with $p_{fa}=0.01$, $K=8$, and $N=20$}
    \label{fig3}
\end{figure}
%-----------------------
\begin{figure}[!tb]
    \centering
    \includegraphics[width =\FW]{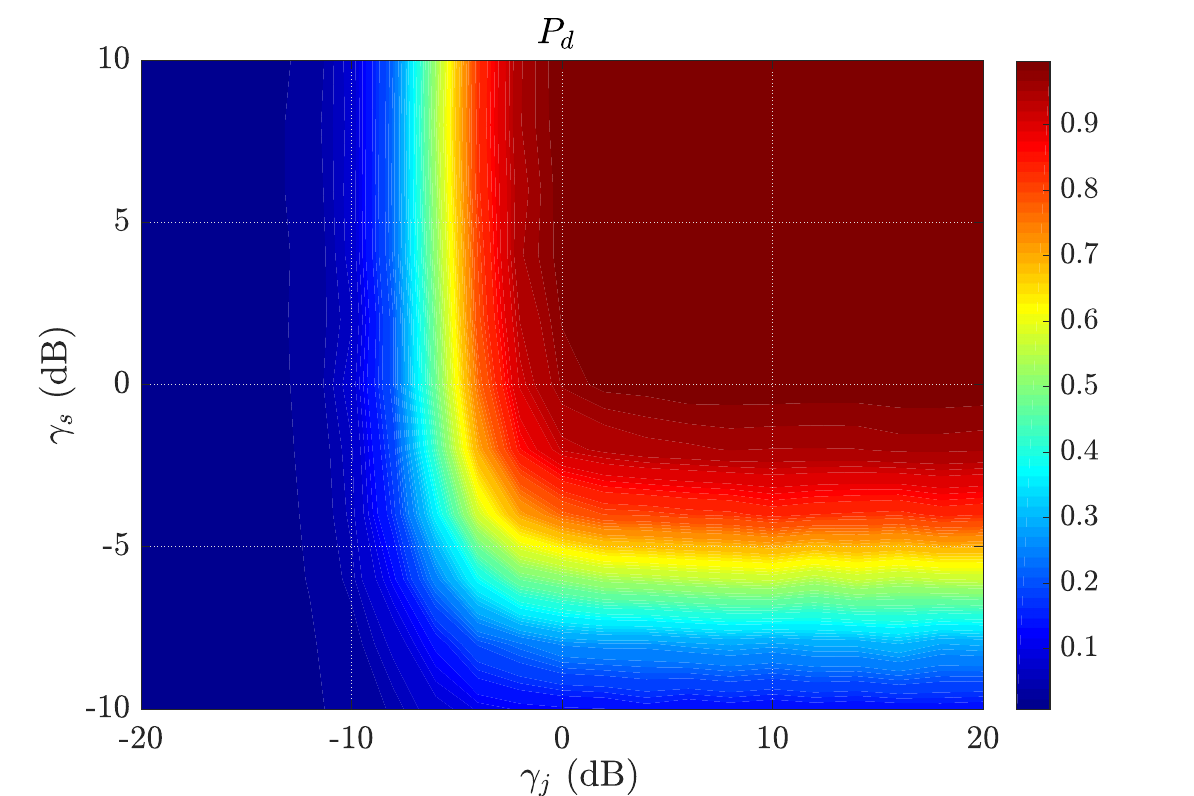}
    \caption{Probability of detection of the RSV detector versus $\gamma_s$ and $\gamma_j$ with $p_{fa}=0.01$, $K=8$, and $N=20$}
    \label{fig4}
\end{figure}
%%%%%%%%%%%%%%%%%%%%%%%%%%%%%%%%%%%%%%%%%%%%%%%%%%%%%%%%%%%%%%%
In Fig. \ref{fig2}, the $P_{fa}$ is computed as a function of the SNR of the desired signals $\hat{\gamma}_s$, while the detection threshold $\eta$ in MC simulations is adjusted based on $\gamma_s=5$ dB. In contrast to the ED, the results of the SSV and RSV detectors imply that both detectors are CFAR with respect to $\gamma_s$ for a quite wide range of discrepancies between $\hat{\gamma}_s$ and $\gamma_s$, which is a satisfactory performance in the context of jamming detection. If $\hat{\gamma}_s$ is extremely small, the SSV and RSV detectors have a smaller $P_{fa}$, which also leads to a decrease in $P_d$ causing misdetection of jamming attacks. To avoid this decrease in the detection performance, using a greater number of SNs is effective. As observed in Fig. \ref{fig2}, for a larger number of received samples $N=50$ and SNs $K=24$, the robustness of the detectors is improved and a decrease in $P_{fa}$ can be observed, even for smaller values of $\hat{\gamma}_s$, in comparison to the case where $K=8$. \par
Furthermore, it is important to note that the KSV and SSV detectors were derived under the assumption of known noise variance, whereas the RSV and GRSV detectors were derived  assuming the noise variance is unknown. Consequently, given the similarity in derivation, a similar CFAR property is anticipated for both the KSV and GRSV detectors.\par
Figs. \ref{fig3} and \ref{fig4} show the contours of the detection probability for the SSV and RSV detectors as a function of $\gamma_s$ and $\gamma_j$. The plotted contours for each detector illustrate the entire detection performance of the SSV and RSV detectors. It is observed that the proposed SSV detector has a superior performance in comparison to the RSV detector since for smaller values of $\gamma_s$ and $\gamma_j$, it achieves higher probabilities of detection and results in a wider red area. For both detectors, it is extremely difficult to detect jammer signals with $\gamma_j<-10$ dB regardless of the value of $\gamma_s$.

\subsection{Comparison with Existing Methods}
\begin{figure}
    \centering
    \includegraphics[width =\FW, height=2.9
    in]{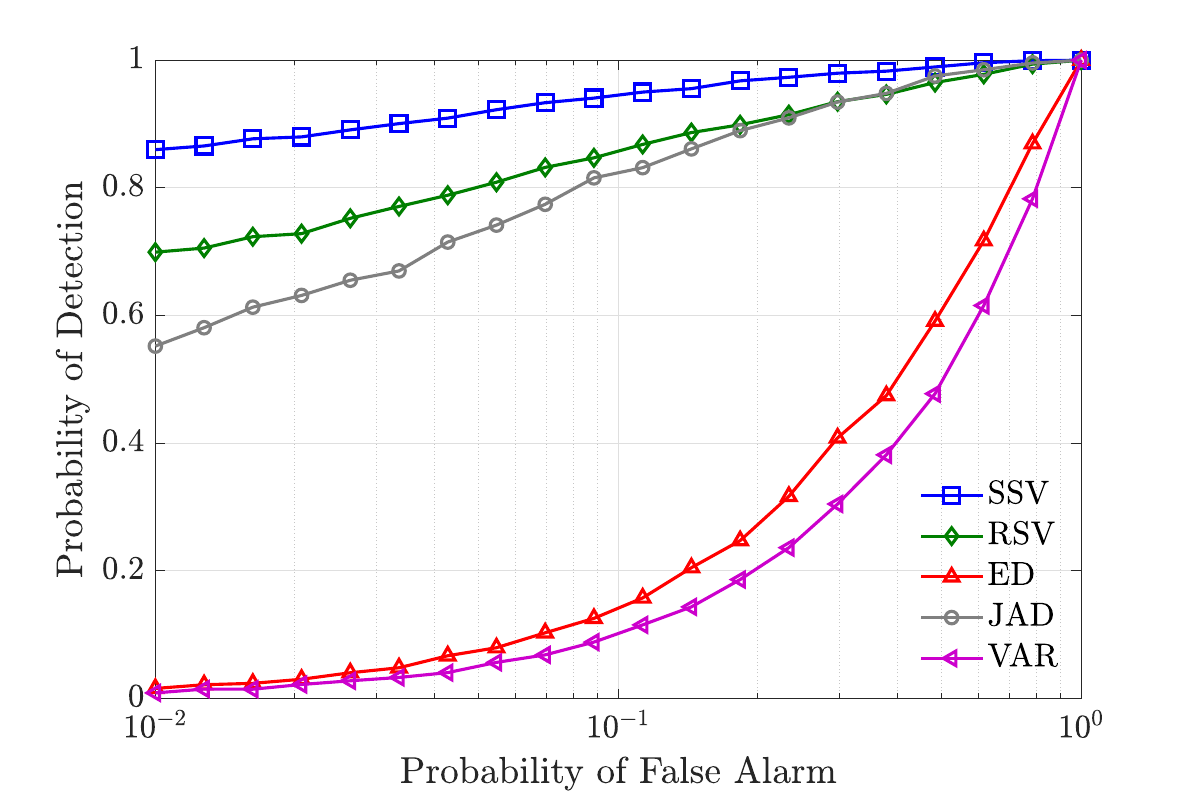}
    \caption{ROC curves for $\gamma_s=5$ dB and $\gamma_j=-5$ dB with $K=8$ and $N=20$}
    \label{fig5}
\end{figure}
Here, we regard the probability of detection for several existing detectors in the literature, such as the jamming attack detector (JAD) in \cite{JAD}, the variance-based detector (VAR) in \cite{VAR}, the number of signal sources detectors according to the random matrix theory (RMT) estimator in \cite{RMT_tugnait, nadler}, the minimum length description (MDL) metric in \cite{MDL}, and Akaike information criterion (AIC) in \cite{AIC}. In systems with multiple antennas, DoFs are directly related to the number of antennas. To ensure a fair comparison of different detection methods, we used the same number of DoFs, i.e.,  $K$, for all detectors, whether they employ multiple antenna systems or multiple sensing nodes, like the proposed detection method. This consistent use of DoFs ensures that all methods are evaluated on an equal basis. %It is very important to note that the number of DoFs ($K$) for these methods with system models with multiple antennas can be interpreted as the number of antennas. Thus, the number of DoFs for all the detectors is the same to compare all methods in an equal setup.  
Additionally, since the AIC and MDL detectors have a fixed detection threshold of one and have no control over the probability of false alarm, we omit them from the receiver operating characteristic (ROC) simulations. We excluded the ROC curve of the RMT and set $\eta$ for the RMT using MC simulations to avoid performance loss. The method in \cite{nadler} might set an inaccurate threshold for the RMT due to the TW distribution approximation, particularly when the necessary condition for the accuracy of TW approximation, which is sufficiently large matrix dimensions,  does not hold. The proposed detector of \cite{Cooperative_WED,kurt} also reduces to the ED for uncorrelated channels which is the case assumed in this paper. Thus, we also simulate the performance of the ED for comparison as well.\par
\begin{figure}
    \centering
    \includegraphics[width =\FW]{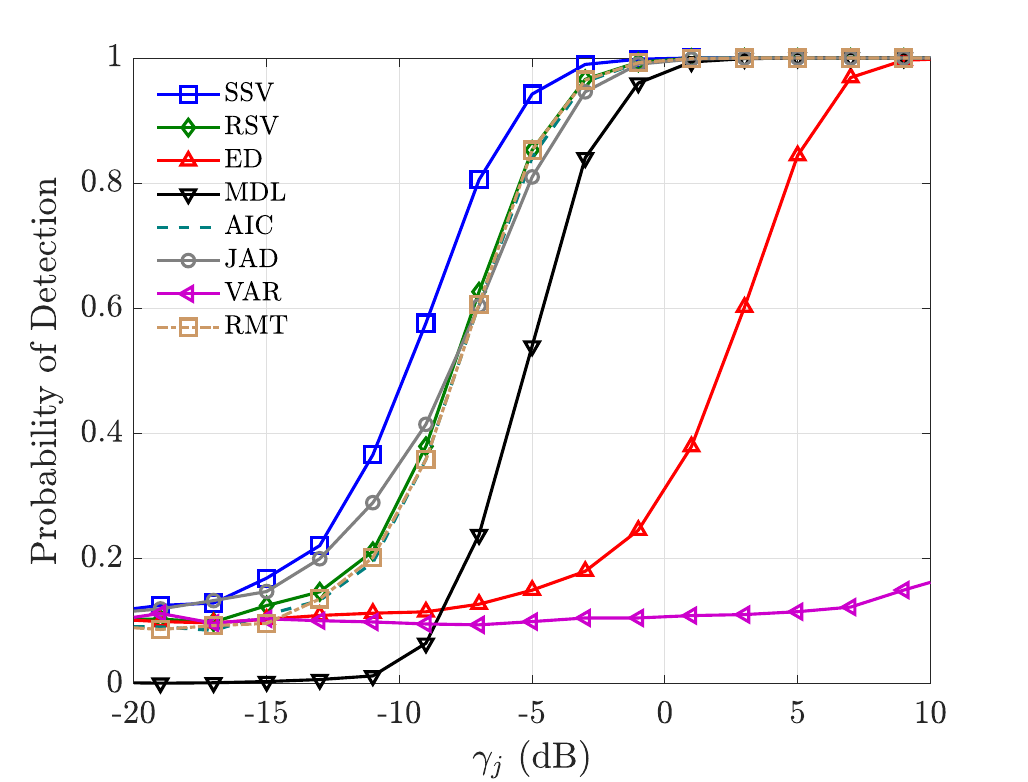}
    \caption{Probability of detection versus $\gamma_j$ with $p_{fa}=0.1$, $K=8$, $N=20$, and $\gamma_s=5$ dB}
    \label{fig6}
\end{figure}
Fig. \ref{fig5} demonstrates the ROC curves of these various detectors where the proposed SSV and RSV detectors outperform the existing detectors like JAD and have a greater area under their ROC curves in comparison with other detectors. For instance, for $P_{fa}=0.01$, a distinct gap exists between the SSV and JAD curves where the SSV detector achieves $P_d=0.85$, while the JAD results in $P_d=0.57$. In addition, other detectors, such as the ED and VAR detectors perform very poorly in this simulation. \par

\begin{figure}
    \centering
    \includegraphics[width =\FW, height=2.92in]{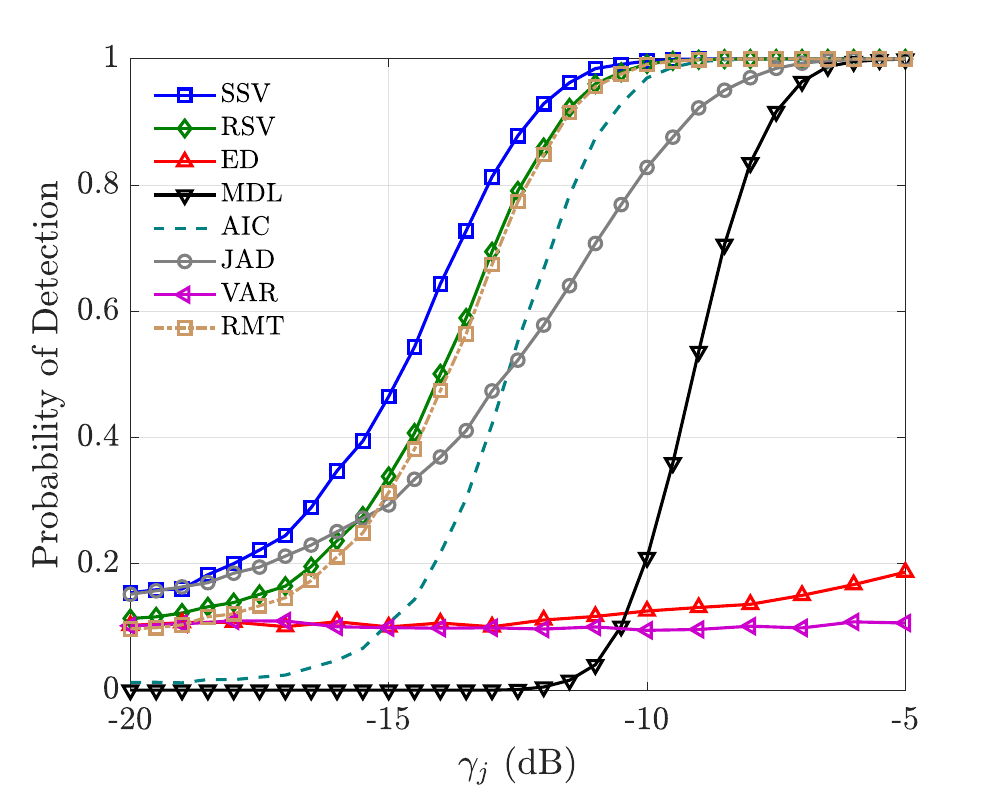}
    \caption{Probability of detection versus $\gamma_j$ with $p_{fa}=0.1$, $K=24$, $N=50$, and $\gamma_s=5$ dB}
    \label{fig7}
\end{figure}

The detection probability as a function of the jammer SNR is depicted in Figs. \ref{fig6} and \ref{fig7}. Inspired by the IEEE 802.22 standard, in these simulations, we have considered an upper bound for false alarm probability as $p_{fa}=0.1$ \cite{IEEE80222}. We can observe that a larger jamming power results in easier detection of the active JN. This figure also confirms the superiority of the SSV detector compared to other detectors. Fig. \ref{fig7} illustrates the same simulation when the number of received samples and SNs are increased. By comparing the results of both figures, we can conclude that a larger number of received samples and SNs improves the performance of all detectors. With a smaller number of SNs, the performance of the JAD,  the RMT, and the proposed RSV detectors are very close while for $K=24$, the RSV and RMT detectors achieve a higher probability of detection than the JAD. Both figures confirm that the SSV detector shows the best performance, and in second rank, the RSV and RMT detectors achieve the highest detection probability. As observed in Fig. \ref{fig7}, to detect a JN with $\gamma_j=-10$ dB, the achieved $P_d$ of the SSV, RSV, and RMT detectors is close to one, while the JAD has a detection probability  of approximately  $0.81$. \par 
%%%%%%%%%%%%%%%%%%%%%%%%%%%%%%%%%%%%%%%%%%%%%%%%%%%%%%%%%%%%%%%%%%%%%%%%%%%%%
\begin{figure}
    \centering
    \includegraphics[width =\FW, height = 2.82 in]{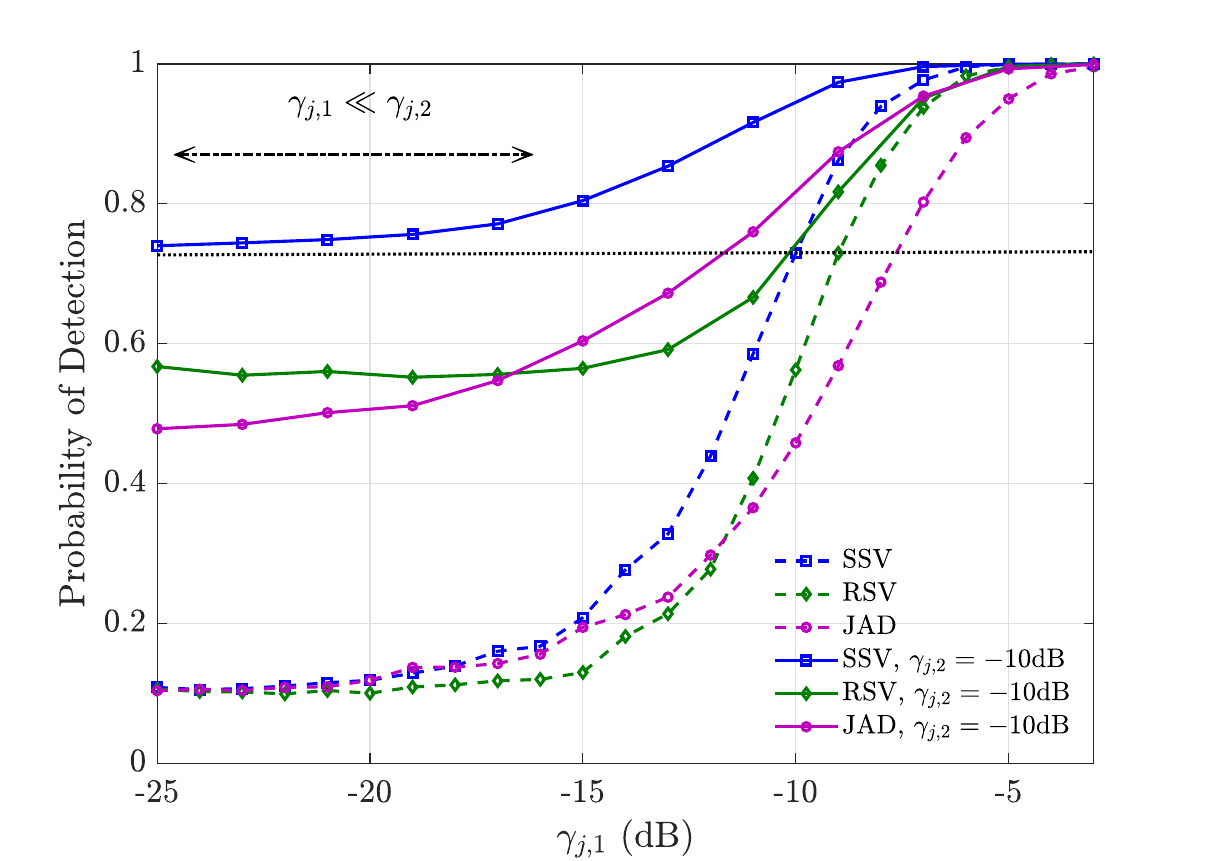}
    \caption{Probability of detection versus $\gamma_{j,1}$ when two JNs exist, where $\gamma_{j,2}=-10$ dB with $p_{fa}=0.1$, $K=16$, $N=20$, and $\gamma_s=5$ dB}
    \label{fig8}
\end{figure}

%%%%%%%%%%%%%%%%%%%%%%%%%%%%%%%%%%%%%%%%%%%%%%%%%%%%%%%%%%%%%%%%%%%%%%%
In  Fig. \ref{fig8}, we add a second JN with SNR equal to $\gamma_{j,2}=-10$ dB to investigate the performance of detectors under multiple sources of jamming since the main system model shown in Fig. \ref{fig0} just included one JN, and the proposed detectors were derived accordingly. The SNR of the first JN in this figure is denoted by $\gamma_{j,1}$.
Additionally, the $P_d$ of detectors in the case of only one JN is also depicted in Fig. \ref{fig8} with the dotted lines for comparison with the case of multiple JNs. As seen in this figure, the existence of a second JN helps detectors achieve a higher $P_d$ when the other JN's SNR is small. In other words, for $\gamma_{j,1}\ll \gamma_{j,2}$, detectors mainly detect the second JN with $\gamma_{j,2}=-10$ by a fixed $P_d\approx 0.75$  since in this curve $\gamma_{j,2}$ is fixed while $\gamma_{j,1}$ changes from $-25$ dB to $-2$ dB. It is noted that $P_d\approx 0.75$ is roughly equal to $P_d$ in the case of a single JN with $\gamma_{j,1}=-10$ dB. This is better shown for the SSV detector by the horizontal dotted black line where this line crosses the curve of the SSV detector with a single JN at $\gamma_{j,1}=-10$ dB. The line roughly touches the SSV detector's curve for two JNs on the left of this figure, where the signal of the first JN is very small, and the SNR of the second JN is $\gamma_{j,2}=-10$ dB. We can conclude that in addition to the superior performance, the suggested detectors can manage not only a situation involving a single JN but also achieve an enhanced detection performance in the case where multiple JNs are active.\par
\begin{figure}
    \centering
    \includegraphics[width =\FW]{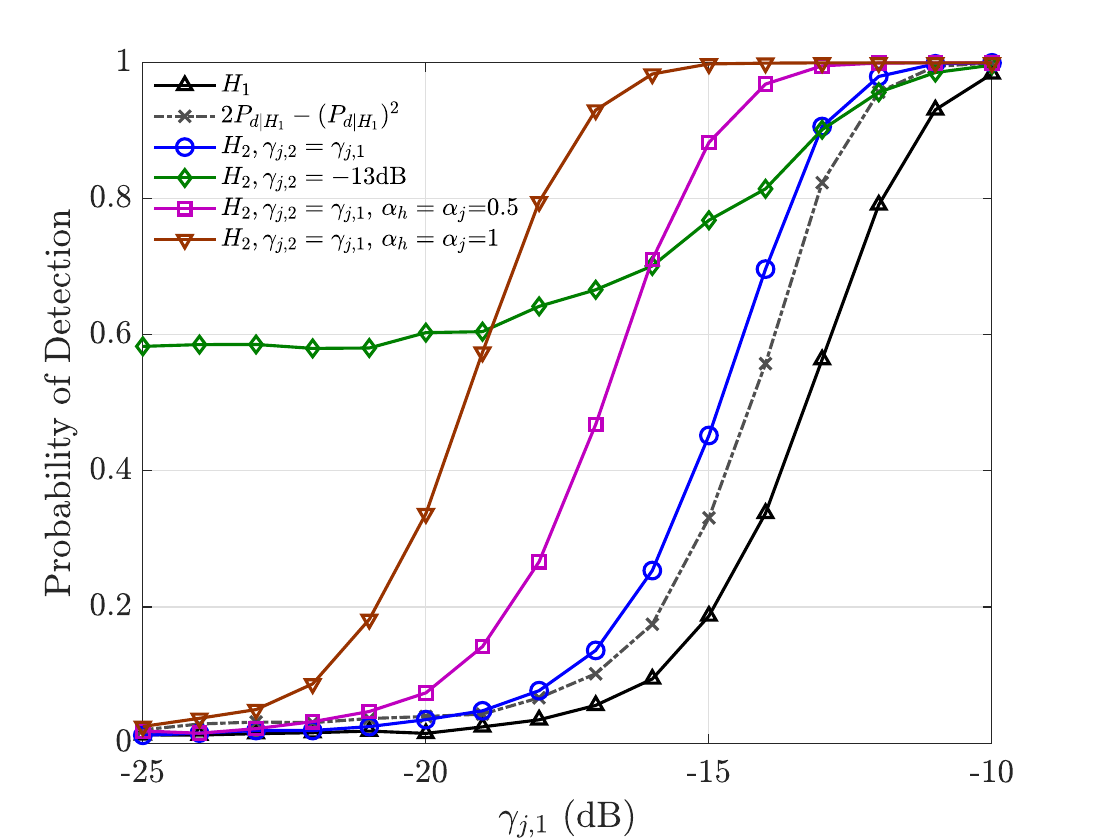}
    \caption{Probability of detection versus $\gamma_{j,1}$ for the SSV detector under $H_1$ and $H_2$ with $K=24$, $N=50$, $\gamma_s=5$ dB, and $p_{fa}=0.01$}
    \label{fig_NEW_MJN}
\end{figure}

For further investigation, Fig. \ref{fig_NEW_MJN}  depicts simulations of the detection probability of the SSV versus $\gamma_{j,1}$ for multiple cases, where $K=24$, $N=50$, $p_{fa}=0.01$, and $\gamma_s=5$ dB. The black curve depicts the SSV results in the presence of only one JN ($H_1$) as a baseline.
The blue curve illustrates $P_d$ of the SSV when two JNs are active ($H_2$), and both have the same transmit SNR, $\gamma_{j,1}=\gamma_{j,2}$. It is clear that the presence of JN2 improves $P_d$  under $H_2$ compared to the case of $H_1$. It is noted that in this scenario, all the channel vectors and symbol vectors are statistically independent, and they are not orthogonal to each other. The dashed curve also shows  $2P_{d|H_1}-(P_{d|H_1})^2$  based on equation (\ref{eq6mjn}) (we omitted superscript ``\textit{o}'' since the orthogonality assumption is discarded). We observe that the SSV under $H_2$ is superior to the dashed curve. In addition, we depict the SSV's $P_d$  when the two JNs have correlated symbol and channel vectors.  We consider $\alpha_j = \alpha_h=0.5$  and  $\alpha_j = \alpha_h=1$, where $\alpha_j$ denotes the correlation between $\textbf{j}_1$ and $\textbf{j}_2$, and  $\alpha_h$ is the correlation coefficient between $\textbf{h}_{j,1}$ and $\textbf{h}_{j,2}$. For both cases, we clearly observe the outperformance of the SSV detector under $H_2$ compared to $H_1$.  Interestingly,  the case of complete correlation ($\alpha_j = \alpha_h=1$) under $H_2$ shows about $6$ dB improvement compared to $H_1$ as we could predict it since the two signals are added, and the amplitude is doubled. The correlated cases can be similar to the case of a JN with multiple antennas, where a correlation between channel vectors of the different antennas can exist due to their close locations. We have also depicted in green the case where JN2 has a constant transmit SNR, i.e.,  $\gamma_{j,2}=-13$ dB. This curve crosses the blue curve at $\gamma_{j,1}=-13$ since at this point, the SNR transmitted by JN1 and JN2 is equal to $-13$ dB for both curves.\par
\begin{figure}
    \centering
    \includegraphics[width =\FW]{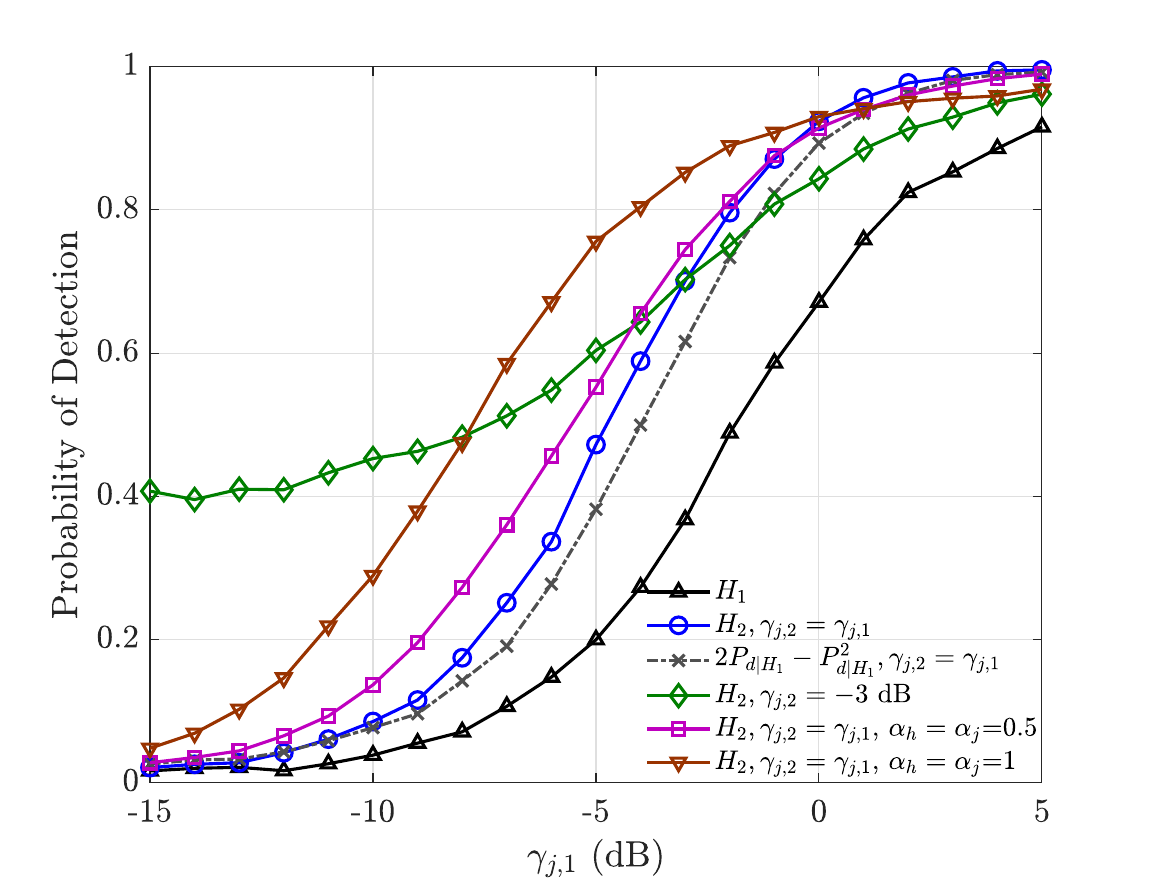}
    \caption{Probability of detection versus $\gamma_{j,1}$ for the SSV detector under $H_1$ and $H_2$ with $K=4$, $N=8$, $\gamma_s=5$ dB, and $p_{fa}=0.01$}
    \label{fig_NEW_MJN2}
\end{figure}
 Fig. \ref{fig_NEW_MJN2} depicts a scenario similar to Fig. \ref{fig_NEW_MJN}  where the number of samples and SNs are reduced to $N=8$ and $K=4$, with $\gamma_{s_t} = 5$ \,dB. We observe that in this scenario, the SSV also attains a higher $P_d$ under $H_2$, whether or not there is a correlation between the JNs' signal and channel vectors. In the case of no correlation, under two JNs, the SSV achieves $P_d = 0.6$ under $H_2$ at approximately $\gamma_{j_1} = -4$ \,dB, while under $H_1$, it requires $\gamma_{j_1} = -1$, dB to achieve the same $P_d$. Even with correlation, we still observe a significant improvement in the probability of detection under $H_2$ compared to $H_1$. As a result, across all these scenarios, it is evident that the proposed detectors become more effective and accurate at detecting jamming attacks when the number of active JNs is increased.

\subsection{Multiple Number of TNs}
In this subsection, we specifically concentrate on the performance of the proposed KSV and GRSV detectors for the scenarios with two TNs, i.e., $M=2$. Here, we consider two TNs with similar transmitting power and SNR  $\gamma_s=5$ dB. We also included the performance of the MDL and AIC as both these methods are based on source separation and have the capability to handle the detection problem for multiple sources of signal. \par
%%%%%%%%%%%%%%%%%%%%%%%%%%%%%%%%%%%%%%%%%%%%%%%%%%%%%%%%%%%%%%%%%%%%%%%%%%%%%
\begin{figure}
    \centering
    \includegraphics[width =\FW, height = 2.87 in]{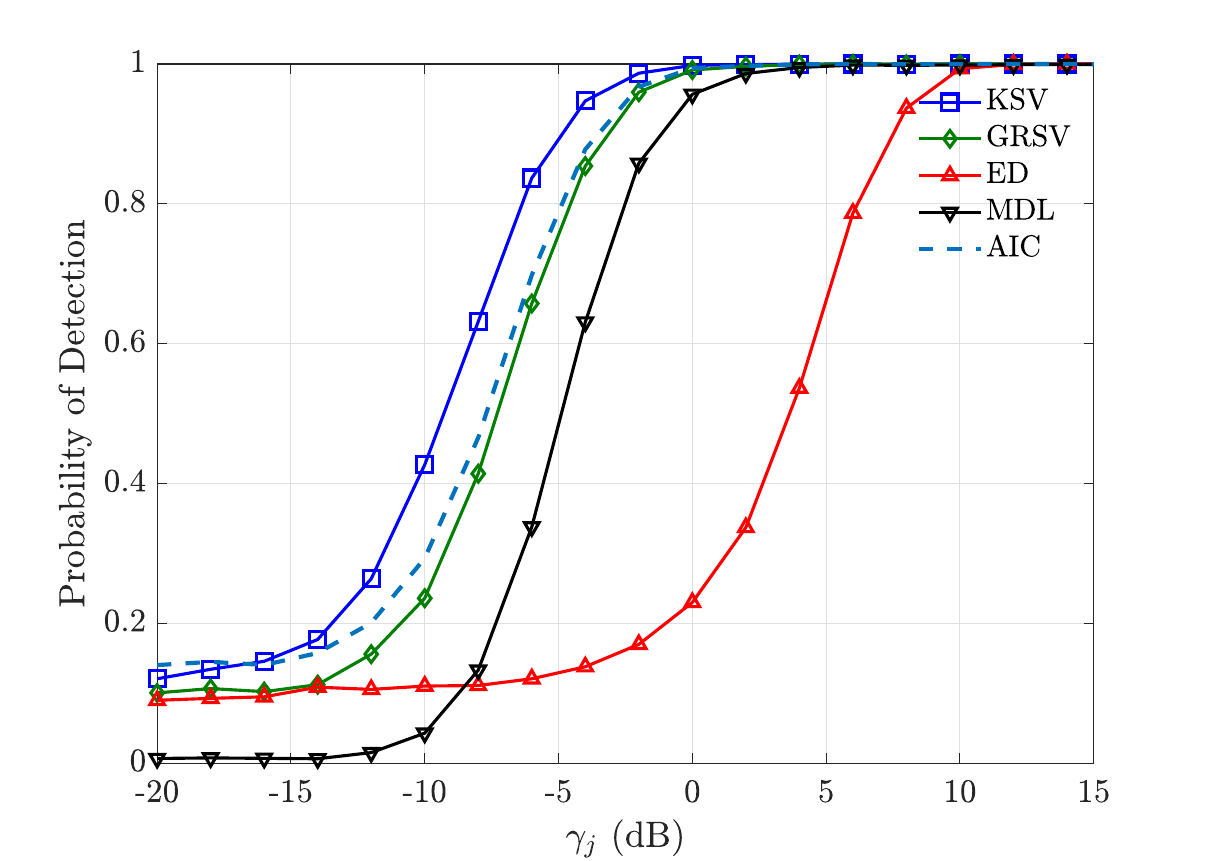}
    \caption{Probability of detection versus $\gamma_j$ with $M=2$, $p_{fa}=0.1$, $K=8$, $N=20$, and $\gamma_s=5$ dB}
    \label{fig12}
\end{figure}
%%%%%%%%%%%%%%%%%%%%%%%%%%%%
Fig. \ref{fig12} shows the probability of detection for the KSV and GRSV detectors as a function of the SNR of the single active JN. Additionally, the number of SNs is $K=8$, and two 
TNs are active. Detectors employ $N=20$ samples for the detection of jamming signals, and the desired probability of false alarm is adjusted to $p_{fa}=0.1$. It is observed that the KSV detector offers the best performance. Following the KSV detector, the AIC exhibits the best performance, closely followed by the performance of the GRSV detector. Specifically, for $\gamma_j=-8$ dB, $P_d$ for the KSV detector is about 20\% higher than that of the AIC detector, highlighting the superior performance of the KSV detector.\par
\begin{figure}
    \centering
    \includegraphics[width =\FW,  height = 2.9 in]{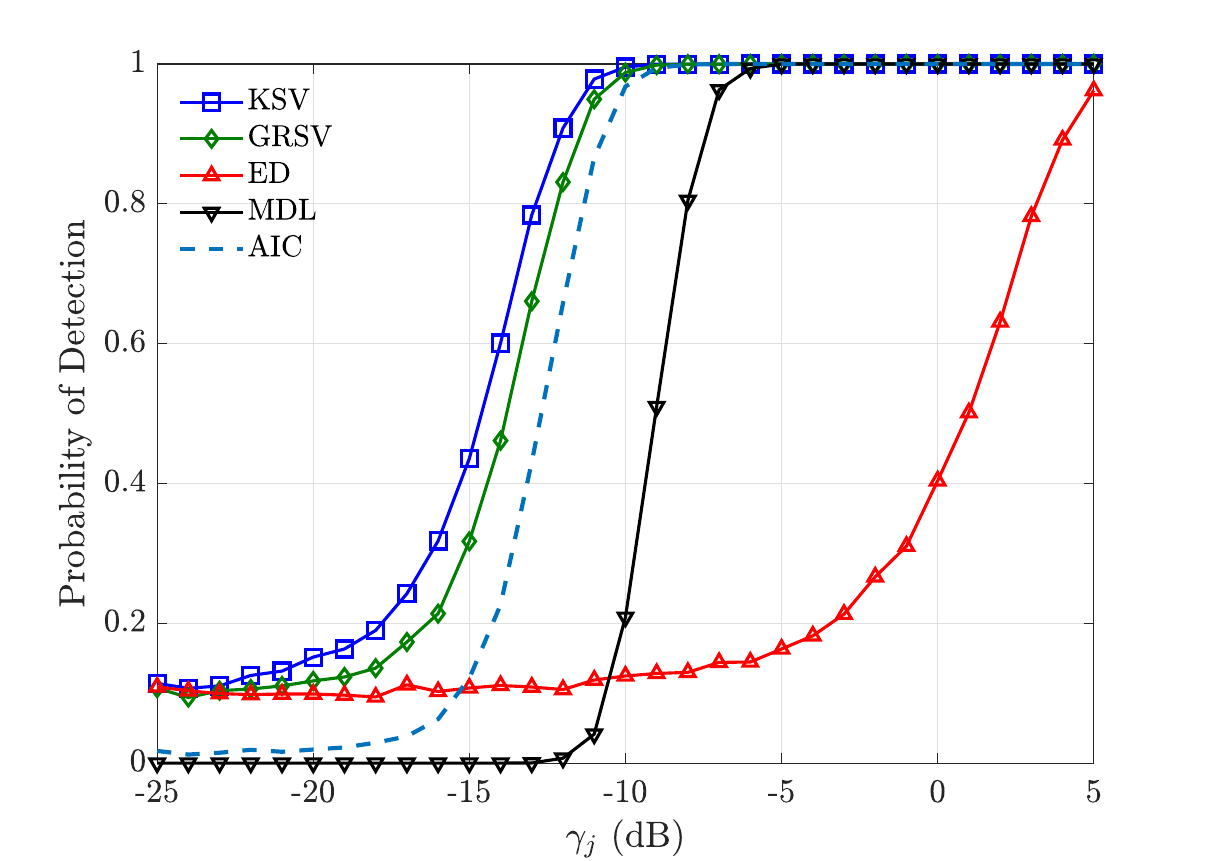}
    \caption{Probability of detection versus $\gamma_j$ with $M=2$, $p_{fa}=0.1$, $K=24$, $N=50$, and $\gamma_s=5$ dB}
    \label{fig13}
\end{figure}

If the number of SNs and sensing samples is increased to 24 and 50, respectively, the performance of the GRSV detector improves and surpasses that of AIC, as evident from Fig. \ref{fig13}. Once again, the KSV detector achieves the highest $P_d$. For instance, at $\gamma_j=-13$ dB, the KSV detector exhibits an approximately 0.35 higher $P_d$ compared to AIC, demonstrating that with a larger number of SNs and sensing samples, the performance of both KSV and GRSV detectors is significantly enhanced.\par
%%%%%%%%%%%%%%%%%%%%%%%%%%%%%%%%%%%%%%%%%%%%%%%%%%%%%%%%%%%%%%%%%%%%%%%
\begin{figure}
    \centering
    \includegraphics[width =3.7in,  height = 2.9 in]{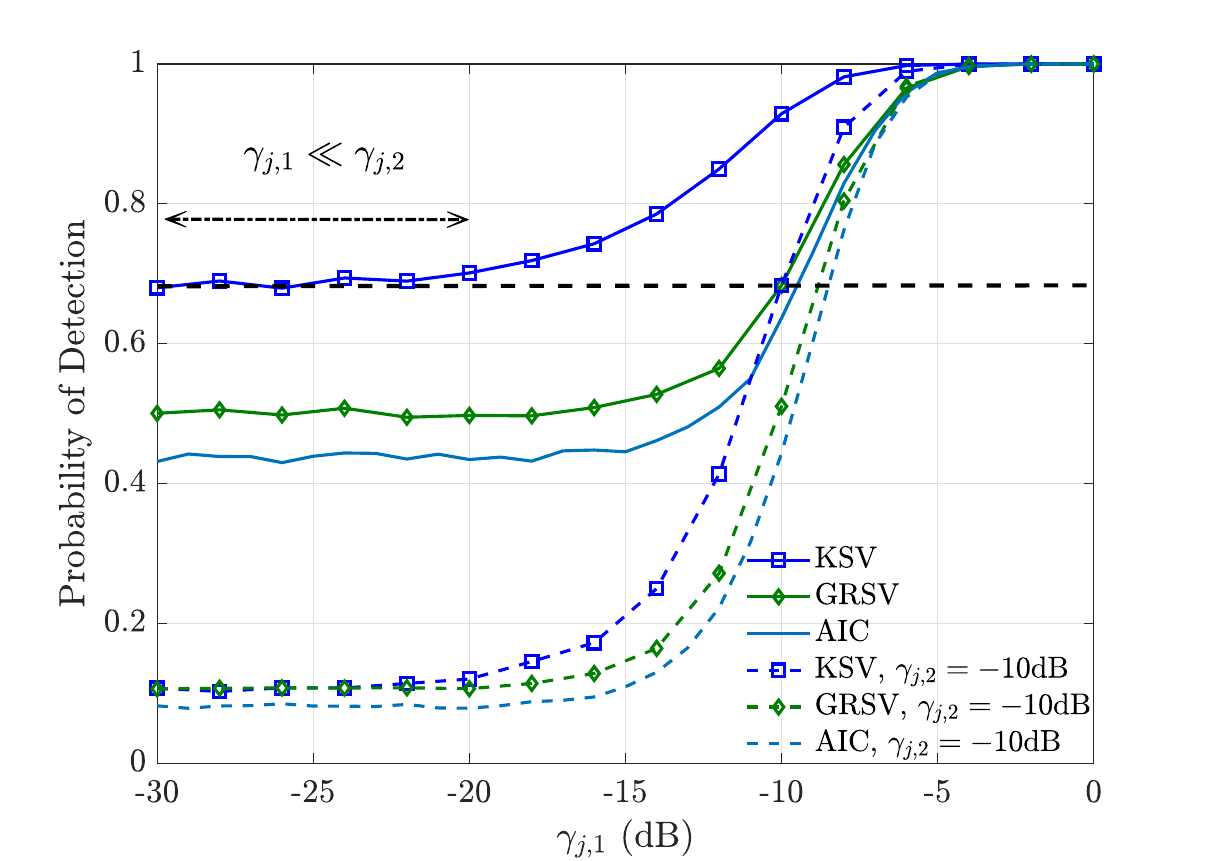}
    \caption{ Probability of detection for versus $\gamma_{j,1}$ when two JNs exist, and for the second JN, $\gamma_{j,2}=-10$ dB with $M=2$, $p_{fa}=0.1$, $K=16$, $N=20$, and $\gamma_s=5$ dB}
    \label{fig14}
\end{figure}
In Fig. \ref{fig14}, we explore a general case involving multiple active TNs and JNs. Specifically, two TNs and two JNs are active, with  $K=16$ and $N=20$. The first JN has a varying SNR denoted by $\gamma_{j,1}$ while the second JN has a fixed SNR, $\gamma_{j,2}=-10$ dB. Additionally, for comparative analysis, this figure incorporates the scenario where only the first JN is active, and the corresponding curves are represented with broken lines. For all three detectors presented in this figure, it is evident that the presence of the second JN helps in detecting the jamming attack, even when the signal from the first JN is very weak ($\gamma_{j,1} \ll \gamma_{j,2} $). When the SNR of the first JN is very small, the detection probabilities of the detectors remain constant and are equal to the case where only one active JN with $\gamma_{j,1}=-10$ dB, as indicated by the black line for the KSV detector.\par
Based on the simulation results of this part, we conclude that the proposed KSV and GRSV detectors, exploiting the low-rank structure of the received matrix, can accurately detect the deceptive jamming attack under multiple numbers of active TNs and JNs.

\section{conclusion}\label{Scon}
In this paper, we introduced effective jamming detection methods for cooperative wireless networks, exploiting the low-rank structure of the received signal matrix, and likelihood ratio tests. The proposed detectors, based on singular value analysis of the received signal matrix demonstrated high accuracy and robustness in the detection of deceptive jamming attacks for scenarios with different numbers of TNs and JNs. We also discussed analytical formulation for the probability of false alarm of the proposed SSV and KSV detectors, which can be used to set the detection threshold. Finally, simulation results and comparisons confirmed the superiority of the proposed methods over several existing techniques, highlighting their potential for enhancing jamming attack detection as the first step for the protection of wireless communication systems against jamming threats. 
 %%also shows under noise uncertainty both ED and sed
%show degradation in performance while unknown sed works well 
 %also regard the point of ED with deceptive jamming where ED works very bad
%also compared with Karblut Kurt method you can compare it with MDL as well
%and a cooperative sensing method
%\\ an ML-based method using SLSV in each node
% ADD general SSV and general RSV
\appendices
\section{LMP test derivation}\label{appA}
To derive the LMP test according to (\ref{eq19}), we start with the LLF of $L(\textbf{Y}| \sigma^2_y)$, which can be stated similar to (\ref{eq6}) as follows
\begin{equation}\label{eqA1}
\begin{split}
L(\textbf{Y}|\sigma^2_y) =  -N\log(\pi)-NK\log(\sigma^2_y)
-(\sigma^2_y)^{-1}\sum_{n=1}^N ||\textbf{y}[n]||_2^2.
\end{split}
\end{equation}
Now, we can find
\begin{equation}\label{eqA2}
\frac{\partial L(\textbf{Y}|\sigma^2_y)}{\partial \sigma^2_y} =  -NK(\sigma^2_y)^{-1}
+(\sigma^2_y)^{-2}\sum_{n=1}^N ||\textbf{y}[n]||_2^2,
\end{equation}
and the second derivative is
\begin{equation}\label{eqA3}
\frac{\partial^2 L(\textbf{Y}|\sigma^2_y)}{\partial (\sigma^2_y)^2} =  NK(\sigma^2_y)^{-2}
-2(\sigma^2_y)^{-3}\sum_{n=1}^N ||\textbf{y}[n]||_2^2.
\end{equation}
The Fisher information can be found using (\ref{eqA4}) \cite{kay-estimation} and the second derivative of the LLF in (\ref{eqA3}) as
\begin{equation}\label{eqA4}
\begin{split}
I&(\sigma^2_y) = \\-&\mathrm{E}\Big[\frac{\partial^2 L(\textbf{Y}|\sigma^2_y)}{\partial (\sigma^2_y)^2} \Big]=-\frac{NK}{(\sigma^2_y)^{2}}
+\frac{2}{(\sigma^2_y)^{3}}\mathrm{E}\Big[\sum_{n=1}^N ||\textbf{y}[n]||_2^2\Big].
\end{split}
\end{equation}
Then, using the fact that $\mathrm{E}\big[\sum_{n=1}^N ||\textbf{y}[n]||_2^2\big]=NK\sigma^2_y$, we have
\begin{equation}\label{eqA5}
\begin{split}
I&(\sigma^2_y) = -\frac{NK}{(\sigma^2_y)^{2}}
+\frac{2NK\sigma^2_y}{(\sigma^2_y)^{3}}=\frac{NK}{(\sigma^2_y)^{2}}.
\end{split}
\end{equation}
Now, using  $\sigma^2_y=\sigma^2_{H_0}$ and plugging (\ref{eqA3}) and (\ref{eqA5}) into (\ref{eq21}), we have
\begin{equation}\label{eqA6}
\begin{split}
T_{LMP}(\textbf{Y})= -\sqrt{NK}
+\frac{1}{\sigma^2_{H_0}\sqrt{NK}}\sum_{n=1}^N ||\textbf{y}[n]||_2^2,
\end{split}
\end{equation}
which is equivalent to the ED.
\bibliographystyle{IEEEtran}
\bibliography{main}

\end{document}